%% file: manuscript.tex
\documentclass[11pt,a4paper]{article}

\usepackage[main=english,croatian]{babel} % for Dokovic2004 (T1 encoding gives bitmap font)
\usepackage[inline]{enumitem}
\setlist[enumerate]{label=(\roman*)}
\usepackage{array}
\usepackage{booktabs}
\usepackage{subcaption}
\usepackage{microtype}

\usepackage{jheppub}
\usepackage{ctns}

\usepackage[noabbrev]{cleveref}
\crefname{equation}{}{}
\Crefname{equation}{Equation}{Equations}
\makeatletter
\AddToHook{cmd/appendix/before}{\def\cref@section@alias{appendix}}
\newcommand\subcapanchors[1]{%
    \begin{subfigure}[b]{0pt}%
        \@for\@subcaplbl:=#1\do{\phantomsubcaption\label{\@subcaplbl}}%
    \end{subfigure}%
}
\makeatother
\makeatletter
\g@addto@macro\@floatboxreset\centering
\makeatother

\RenewDocumentEnvironment{bmatrix}{b}{%
  \left[\,\begin{matrix}#1\end{matrix}\,\right]}{}

\begin{document}

\title{Gaussian continuous tensor network states:\\ short-distance properties and imaginary-time evolution}
\author[a,b]{Marco Rigobello}
\author[c]{Erez Zohar}

\affiliation[a]{Max-Planck-Institut für Quantenoptik, Hans-Kopfermann-Str.~1, D-85748 Garching, Germany}
\affiliation[b]{Dipartimento di Fisica e Astronomia ``G. Galilei'', Università di Padova, I-35131 Padova, Italy}
\affiliation[c]{School of Physics and Astronomy, Tel Aviv University, Tel Aviv 6997801, Israel}

\emailAdd{marco.rigobello@mpq.mpg.de}
\emailAdd{erezzohar@tauex.tau.ac.il}

\abstract{
    We study Gaussian continuous tensor network states (GCTNS) ---
    a finitely parameterized subclass of Gaussian states
    admitting an interpretation as continuum limits of discrete tensor network states.
    We show that, at short distance, GCTNS correspond to vacua of free Lifshitz theories with even dynamical exponent, establishing a connection between certain entanglement properties of the two.
    Two schemes for approximating ground states of free bosonic field theories using GCTNS are presented: rational approximants to the exact dispersion relation and Trotterized imaginary-time evolution.
    We apply them to Klein-Gordon theory and characterize the resulting GCTNS, identifying the scale at which deviations from its conformal short-distance behavior appear.
    In particular, we construct an optimal rational approximation for which, at fixed accuracy, the bond dimension grows only logarithmically with the UV cutoff.
    These results provide a simple and analytically controlled setting to assess the strengths and limitations of GCTNS as variational ansätze for relativistic quantum fields.
}

\arxivnumber{2602.15987}

\maketitle

\section{Introduction}\label{sec_introduction}

Tensor network states (TNS) \cite{Verstraete2008MatrixProductStates,Orus2014PracticalIntroductionTensor} are powerful tools for studying strongly correlated quantum systems.
They have contributed to important analytical insights, such as the classification of symmetry-protected and topological phases of matter \cite{Cirac2021MatrixProductStates}, and have broad numerical applications \cite{Silvi2019TensorNetworksAnthology,Banuls2023TensorNetworkAlgorithms}.
By parametrizing many-body wavefunctions as contractions of low-rank tensors along virtual spaces, TNS capture the entanglement structure of local lattice models with resources that grow only polynomially with the system size,
making them highly effective as variational classes.
The most prominent example is the matrix product state ansatz (MPS), which underlies the density-matrix renormalization group and provides quasi-exact results for spin chains \cite{Cirac2021MatrixProductStates,Schollwoeck2011DensityMatrixRenormalization}.
Projected entangled-pair states (PEPS) \cite{Verstraete2004RenormalizationAlgorithmsQuantum} extend the MPS construction and its area-law entanglement to higher space dimensions; although optimization is much more demanding, PEPS have achieved state-of-the-art results, \eg{} for the doped 2$\spdim$ Hubbard model \cite{Liu2025AccurateSimulationHubbard}.
Over the years, other ansätze have been introduced, such as tree tensor networks (TTN) \cite{Shi2006ClassicalSimulationQuantum} and the multi-scale entanglement renormalization ansatz (MERA) \cite{Vidal2007EntanglementRenormalization}, especially tailored to critical phenomena \cite{Giovannetti2008QuantumMultiscaleEntanglement}.

TNS have long been employed also to study systems with a continuous number of degrees of freedom, described by quantum field theory (QFT).
A prominent focus of this program has been on low-dimensional gauge theories \cite{Banuls2013MassSpectrumSchwinger}, where TNS algorithms have enabled non-perturbative investigations of problems intractable for sampling methods \cite{Troyer2005ComputationalComplexityFundamental}, such as finite fermion density and out-of-equilibrium scenarios
\cite{Banuls2020SimulatingLatticeGauge,Kelman2024GaugedGaussianProjected,Magnifico2025TensorNetworksLattice}.
As in Monte Carlo simulations \cite{Duerr2008AbInitioDetermination},
this is traditionally achieved by first discretizing the theory on the lattice, thereby introducing a short-distance cutoff,
and finally extrapolating results to the continuum \cite{Kogut1983LatticeGaugeTheory}.
It is nonetheless possible to formulate TNS directly in the continuum.
Defining sparse variational manifolds capable of encoding QFTs at all length scales is both conceptually appealing and, potentially, computationally relevant, as it avoids delicate extrapolations in the lattice spacing.
The breakthrough in this direction came with the introduction of continuous matrix product states (CMPS) \cite{Verstraete2010ContinuousMatrixProduct}, which realize a genuine continuum limit of MPS \cite{DelasCuevas2018ContinuumLimitsMatrix}
and allow for the efficient computation of observables and variational gradients in 1$\spdim$ \cite{Haegeman2013CalculusContinuousMatrix,Ganahl2017ContinuousMatrixProduct,Tuybens2022VariationalOptimizationContinuous}.
Generalization to higher space dimensions required substantial efforts and led to continuous versions of PEPS \cite{Jennings2015ContinuumTensorNetwork} and MERA \cite{Haegeman2013EntanglementRenormalizationQuantum}.
All these were later unified under the umbrella of continuous tensor network states (CTNS) \cite{Tilloy2019ContinuousTensorNetwork}, clarifying their origin as continuum limits of discrete TNS with the correct spatial symmetries \cite{Tilloy2019ContinuousTensorNetwork,Shachar2022ApproximatingRelativisticQuantum}.
This limit has been recently constructed also for gauged PEPS \cite{Roose2025ContinuumLimitGauged}, which span the Hilbert space of lattice gauge theories \cite{Blanik2025InternalStructureGauge}.
Crucially, while the one-dimensional CMPS retain a finite virtual space, consistency of the continuum limit in $\spdim\mathbin>1$ demands infinite-dimensional virtual spaces \cite{Tilloy2019ContinuousTensorNetwork},
making the study of general CTNS virtually as challenging as the original QFT problems they were meant to solve.
The Gaussian subclass (GCTNS), however, is analytically tractable by leveraging Wick's theorem, and faithful descriptions of quadratic theories, as well as controlled approximations at weak coupling, have been worked out explicitly \cite{Karanikolaou2021GaussianContinuousTensor,Shachar2022ApproximatingRelativisticQuantum}.

In this work we analyze structural properties of Gaussian CTNS, focusing especially on their short-distance behavior.
The main motivation for this analysis comes from relativistic QFT.
In the relativistic setting, CMPS and CTNS are known to require an ultraviolet (UV) regulator to stabilize variational optimizations \cite{Haegeman2010ApplyingVariationalPrinciple,Stojevic2015ConformalDataFinite,Tilloy2021RelativisticContinuousMatrix,Karanikolaou2021GaussianContinuousTensor}, reflecting their mismatch with the conformal short-distance behavior of relativistic QFTs, and reintroducing the need for extrapolations.
In $1\spdim$, a relativistic modification of CMPS (RCMPS) remedies this by aligning the ansatz with the correct
free-theory Fock structure
\cite{Tilloy2021VariationalMethodRelativistic,Tilloy2021RelativisticContinuousMatrix}.
In higher-dimensional interacting theories, however, divergences are structurally more severe and optimization will still require explicit cutoffs.
In this regard, the Gaussian sector provides a valuable controlled setting to assess the strengths and limitations of continuous tensor network states beyond one space dimension \cite{Karanikolaou2021GaussianContinuousTensor}.

The manuscript is organized as follows:
\Cref{sec_ctns} sets the notation by recalling the definition of CTNS and GCTNS, and briefly discusses the parent Hamiltonians and entanglement properties of the latter.
In \cref{sec_gs} we analytically construct approximate GCTNS ground states, thereby showing how Trotterized imaginary-time evolution can be interpreted as a CTNS.
\Cref{sec_kg} benchmarks these approximations against a free relativistic QFT, inspecting their UV properties.
\Cref{sec_conclusion} summarizes our findings and discusses their implications, comparing to non-relativistic settings, CMERA and RCMPS.

\section{\CTNS{} and Gaussian \CTNS{}}\label{sec_ctns}

Consider a real
scalar field operator $\ffop$ supported on $\spdim$-dimensional space or, in the relativistic case, on a constant-time slice of ($\spdim$+1)-dimensional Minkowski spacetime.
Introducing a basis of field configuration eigenstates,
\begin{equation*}
    \ffop(\xxx)\ket{\ff} = \ff(\xxx)\ket{\ff}
    \ ,
\end{equation*}
any state $\ctns>$ of $\ffop$ may be identified with the Schrödinger wavefunctional
\begin{math}
    \Psi\colon \ff \mapsto \braket{\ff}{\Psi}
\end{math}.
A CTNS or CPEPS \cite{Tilloy2019ContinuousTensorNetwork,Shachar2022ApproximatingRelativisticQuantum}
for $\ffop$ is a state whose wavefunctional is expressed as a path integral over $\cbd$ \emph{auxiliary} or \emph{virtual} fields
\cite{Shachar2022ApproximatingRelativisticQuantum}\footnote{
    The equivalence of this definition to that of ref.~\cite{Tilloy2019ContinuousTensorNetwork} was proven in ref.~\cite[appendix~A]{Shachar2022ApproximatingRelativisticQuantum}.
    Generalization to curved spaces or systems with a boundary is possible
    \cite{Tilloy2019ContinuousTensorNetwork}.
    In this work we only consider field theories on Euclidean $\Gali$
    and use the name CPEPS to stress this fact.
}
\begin{align}\label{eq_ctns}
    \ctns(\ff) = \int\DD\ffv_1\cdots\DD\ffv[-1] \: e^{-\ctnsfunc(\ff,\ffv_1,\dots,\ffv[-1];\ctnspar)}
    \ ,
\end{align}
where $\ctnsfunc$ is a \emph{local} functional of the fields
depending on a \emph{finite} set of parameters $\ctnspar$,
whose number, together with $\cbd$, controls the expressivity of the CTNS class.\footnote{%
    That is, $\ctnsfunc$
    is the space integral of a polynomial of a finite number of derivatives of the fields and functions of the field values, all evaluated at the same position.
    In order to get a finite number of parameters $\ctnspar$, functions of the fields may be expanded in some finite basis set, \eg{} a monomial basis \cite{Karanikolaou2021GaussianContinuousTensor}.
}
We further require $\ctnsfunc$ to be invariant under space rotations and translations.
It has been shown that, with these assumptions,
$\ctns$
in \cref{eq_ctns}
is the continuum limit of an ordinary PEPS whose auxiliary spaces correspond each to the (infinite-dimensional) Fock space of $\cbd$ bosonic creation operators \cite{Tilloy2019ContinuousTensorNetwork,Shachar2022ApproximatingRelativisticQuantum}.
In this sense, the number of auxiliary fields $\cbd$ may be regarded as the field-theory analogue of the bond dimension.
If the functional $\ctnsfunc'$ obtained integrating one virtual field, say $\ffv[-1]$,
\begin{equation}\label{eq_ctns_reduce}
    \def\vindmax{\cbd-1}
    e^{-\ctnsfunc'(\ff,\ffv_1,\dots,\ffv[-1])}
    \let\vindmax\cbd
    = \int\!\DD\ffv[-1] \: e^{-\ctnsfunc(\ff,\ffv_1,\dots,\ffv[-1])}
    \ ,
\end{equation}
is still local, $\ctnsfunc'$ also defines a \CTNS{} (with $\cbd-1$ virtual fields).
Of course, virtually no closed-form solutions for QFT path integrals such as the above are known, aside from Gaussian integrals and a few other very special exceptions.

In the Gaussian case ($\ctnsfunc$ quadratic in the fields),
translation invariance makes it particularly convenient to work in momentum space.
Collecting all the fields in
\begin{math}
    \fff=\{\ffv\}\fvlims
\end{math},
with $\ffv[0]\mathbin=\ff$ the physical field,
a G\CTNS{} with all the desired properties is obtained setting
\cite{Shachar2022ApproximatingRelativisticQuantum,Karanikolaou2021GaussianContinuousTensor}\footnote{%
    Fields are always assumed to satisfy suitable boundary conditions so that total derivatives can be discarded when integrating by parts.
}
\begin{equation}\label{eq_gctns_W}
    \ctnsfunc(\fff)
    = \frac{1}{2}\xint\:\fff(\xxx)\transp\brackets[\big]{\ctnsmat{0}-\ctnsmat{1}\nabla^2}\fff(\xxx)
    = \frac{1}{2}\kint\:\fff(-\kkk)\transp\ctnsmat(\kk)\fff(\kkk)
    \ ,
\end{equation}
where
$\ctnsmat{0}, \ctnsmat{1}\in\mathcal{M}_{\cbd+1}(\complex)$
are constant symmetric matrices
(the ansatz parameters $\ctnspar$),
and
\begin{math}
    \ctnsmat(\kk) \mathbin= \ctnsmat{0} + \ctnsmat{1}\kk^2
\end{math} with
$\kk \mathbin= \norm{\kkk}$.
Fields and their Fourier transforms are denoted in the same way and we often omit their $\xxx$ or $\kkk$ argument for brevity
(they can be discerned from the integration measure),
in which case $\ff*$ denotes $\ff(-\kkk)$.
In principle $\ctnsfunc$ could include terms constant and linear in $\fff$; however, the former only affect the normalization and global phase of $\ctns$, which we are not going to keep track of, and
linear terms can be reabsorbed by a constant shift of the fields.
Clearly, the shift has physical consequences if $\ffv[0]$ is involved;
here we assumed $\vev{\ffop}^{}_{\ctns}=0$.
Finally, higher-order derivatives of $\fff$ (\idest{}, letting $\ctnsmat$ be a finite polynomial of $\kk^2$) can be traded for additional virtual fields.
More details on the general structure of the $\ctnsfunc$ functional can be found in \cref{sec_canonical,sec_higher_order}.
Unless otherwise stated, here we assume it takes the form \cref{eq_gctns_W}.

\emph{Contracting} the G\CTNS{} --- namely, integrating over the virtual fields --- yields, up to
normalization, a Gaussian state
\begin{equation}\label{eq_gctns_contracted}
    \ctns(\ff) = \exp*{-\frac{1}{2}\kint\, \kernel(\kk)\ff*\ff}
    \ ,
\end{equation}
whose Gaussian kernel $\kernel$ is the Schur complement of the virtual ($\fvind|,|[1;2]>\fvind[0]$) block $\vctnsmat[1;2](\kk)$ in $\ctnsmat[1;2](\kk)$ \cite{Shachar2022ApproximatingRelativisticQuantum}:
\begin{equation}\label{eq_gctns_kernel}
    \kernel
    = \ctnsmat[0;0]
    - \sum\vvlims[1;2] \ctnsmat[0;1][\vctnsprop]\indices*{_{\vvind[1;2]}}\ctnsmat[2;0]
    \ .
\end{equation}
The second term captures the contribution from the coupling of $\ff$ to the integrated-out virtual fields, whose propagator $\vctnsprop$ can be written in terms of the adjugate matrix
as
\begin{equation*}
    \vctnsprop = \frac{\adj\vctnsmat}{\det\vctnsmat}
    \ .
\end{equation*}
Since the entries of $\ctnsmat$ are polynomials in $\kk^2$,
one concludes that $\dvord$th-derivative G\CTNS{} with $\cbd$ auxiliary fields
are Gaussian states whose kernel $\kernel$ is a \emph{rational function of $\kk^2$} of degree $\dvord(\cbd+1)$ over $\dvord\cbd$ \cite{Shachar2022ApproximatingRelativisticQuantum}.
Conversely, as shown in \cref{sec_canonical}, any $\kernel$ rational in $\kk^2$ is realized by G\CTNS{} --- provided it has no poles at physical momenta, $\kk^2\geq0$, and $\re\kernel>0$ there, so that \cref{eq_gctns_contracted} defines a normalizable Gaussian state.\footnote{%
    Poles of $\kernel$ and zeros of $\re\kernel$ at $\kk^2\geq0$ can arise in infinitely squeezed limits of normalizable G\CTNS{}.
}
This rational kernel property mirrors a similar result for (discrete) Gaussian MPS \cite{Schuch2008GaussianMatrixProduct}.
Throughout this work, we will highlight many of its consequences.
An immediate one is that all the correlation functions of $\ffop$ and its conjugate momentum $\ppop$ on $\ctns$ are also rational; \eg{},
\begin{equation}\label{eq_gctns_correlators}
    \vev{\ffop(\kkk[2])\ffop(\kkk[3])} = (2\pi)^\spdim\delta(\kkk|+|[2;3])\frac{1}{2\re\kernel(\kk[2])}
    \ , \quad
    \vev{\ppop(\kkk[2])\ppop(\kkk[3])} = (2\pi)^\spdim\delta(\kkk|+|[2;3])\frac{\abs{\kernel(\kk[2])}^2}{2\re\kernel(\kk[2])}
    \ ,
\end{equation}
and higher-order ones follow from Wick's theorem.

\paragraph{Gauge fixing.}
A second observation is that the rational kernel $\kernel$ (and thus the state $\ctns$) is specified by a number of parameters linear in $\cbd$, while $\ctnsmat$ (and thus the G\CTNS{} ansatz) has $\order{\cbd^2}$ parameters.
This over-parameterization is symptomatic of the reparameterization freedom of the path integral \cref{eq_ctns} defining the G\CTNS{}; it means that different $\ctnsmat$, related by the transformation \cref{eq_gctns_transf_W}, give the same $\kernel$.
Such ``gauge'' freedom is common in TNS and can be removed by bringing the ansatz into \emph{canonical form},
often making numerical optimization better conditioned \cite{Verstraete2008MatrixProductStates}.
For instance,
redundancy introduces flat directions in the variational landscape that can significantly slow down gradient-based optimization.
We introduce below two canonical forms for generic G\CTNS{}, and refer to \cref{sec_canonical} for further details and degenerate cases.

\begingroup
\let\polenum\vvind
\let\poleind\vvind
Consider a diagonal virtual block and non-derivative physical-virtual couplings, setting
\begin{equation}\label{eq_gctns_diagonal}
    \ctnsfunc(\fff)
    =
    \frac{1}{2} \kint \brackets*{
        (a + b\kk^2) \, \ffdia[0]
        + \sum\vvlims (\polemass + \kk^2) \, \ffdia
        + 2i\sum\vvlims \sqrt{\poleres} \, \ffv*[0]\ffv
    }
    \ .
\end{equation}
\Cref{eq_gctns_kernel} then yields the linear partial fraction expansion
\begin{equation}\label{eq_rational_simple_poles}
    \kernel(\kk) =
    a + b\kk^2
    + \sum\vvlims \frac{\poleres}{\polemass + \kk^2}
    \ ,
\end{equation}
showing that it was not restrictive to normalize the virtual fields' kinetic terms ($\vctnsmat{1}\mathbin=\idmat$) in \cref{eq_gctns_diagonal}:
$\vctnsmat[1;1]{1}\mathbin{\neq}1$ can be compensated by rescaling $\poleres$ and $\polemass$, unless $\vctnsmat[1;1]{1}\mathbin=0$, in which case $\ffv$ is spurious as it contributes a constant to $\kernel$, which can be absorbed in $a$.
Without loss of generality, we can further assume
\begin{math}
    \poleres\mathbin{\neq}0
\end{math}
(otherwise $\ffv$ decouples),
and $\polemass[2]\mathbin{\neq}\polemass$ for $\vvind|\mathbin{\neq}|[2;1]$ (otherwise $\ffv[2]$ is redundant).
With these assumptions, any rational $\kernel$ having only simple poles, with numerator of degree at most $\cbd+1$ and denominator of degree exactly $\cbd$, can be uniquely written as \cref{eq_rational_simple_poles} \cite{EMS2025RationalFunction}.
This is, in fact, the \emph{generic} case;
higher-order poles and sub-maximal degree denominators can nonetheless be recovered, \eg{} taking $\vctnsmat$ in \emph{symmetrized Jordan normal form} and introducing some derivative couplings, as in \cref{eq_canonical_jordan}.
\endgroup

Another useful choice, which will naturally appear in the following, is to take $\ctnsmat$ \emph{tridiagonal}.
$\kernel$ is easily computed integrating out virtual layers one by one, as per \cref{eq_ctns_reduce}.\footnote{%
    See also ref.~\cite{Usmani1994InversionJacobisTridiagonal} for a direct computation of $\vctnsprop$.
}
The functional obtained integrating out $\ffv[-1],\dots,\ffv_{\rcbd+1}$ reads
\begin{equation}\label{eq_gctns_tridiagonal}
    \def\vindmax{\rcbd}
    \ctnsfunc^{}_{\!\vindmax}(\ffv[0], \dots, \ffv[-1])
    =
    \frac{1}{2} \kint \brackets*{\,
        \sum\fvlims^{\rcbd-1}\tridia(\kk) \, \ffdia
        + \ctnslea{\vindmax}(\kk) \, \ffdia[-1]
        + 2i\sum\vvlims\trioff(\kk) \, \ffoff
    }
    \,.
\end{equation}
Indeed, $\ctnsfunc_{\!\cbd}=\ctnsfunc$ has this structure by hypothesis, and Gaussian integration
only updates the trailing main-diagonal coefficient:
\begin{equation*}
    \ctnslea{\rcbd} = \tridia_{\rcbd} + \trioff_{\rcbd+1}^2/\ctnslea{\rcbd+1}
    \ , \quad
    \ctnslea{\cbd} = \tridia[-1]
    \ .
\end{equation*}
Iterating until $\rcbd=0$, one gets the (finite) \emph{continued fraction},
\begin{equation}\label{eq_tridiagonal_cf}
    \kernel
    = \ctnslea{0}
    = \tridia_0 + \cfrac{\trioff_1^2}{
        \tridia_1+\cfrac{\trioff_2^2}{
            \cfracdots+\cfrac{\trioff[-1]^2}{\tridia[-1]}}}
    \ .
\end{equation}
Using the Euclidean algorithm (on $\complex[\kk^2]$), \emph{generic} rational functions of degree $\cbd+1$ over $\cbd$ can be written in this way, with $\trioff=1$ \cite{Sauer2021RationalFunctionsAs}.
The convergence of continued fractions has been extensively studied \cite{Lorentzen2008ContinuedFractionsConvergence}, making this representation particularly convenient when discussing approximations of non-rational kernels $\kernel$ by means of GCPEPS, in \cref{sec_gs}.

\subsection{Parent Hamiltonians and renormalization}\label{sec_parents}
Key properties of TNS can be understood through their parent Hamiltonians, namely (quasi\nobreakdash-)\allowbreak local Hamiltonians that have a given TNS as their ground state \cite{Cirac2021MatrixProductStates}.
$\ctns$ in \cref{eq_gctns_contracted} is the unique ground state of a family of quadratic Hamiltonians,
parameterized by their (arbitrary) dispersion relation $\freq$,
\begin{equation}\label{eq_parent}
    \pham = \frac{1}{2}\kint \frac{\freq(\kk)}{\re\kernel(\kk)}\brackets[\Big]{
        \ppop*\ppop
        +\abs{\kernel(\kk)}^2\ffop*\ffop
        +\im\kernel(\kk)\brackets[\big]{\ppop*\ffop + \ffop*\ppop}
    }
    \ .
\end{equation}
For simplicity, hereafter we assume time-reversal symmetry, namely $\im\kernel=0$.
Choosing $\freq$ appropriately, $\pham$ can be taken local \cite{Shachar2022ApproximatingRelativisticQuantum}. Instead,
canonically normalizing the conjugate momentum by setting $\freq=\kernel$, the class of G\CTNS{} is seen to encode \emph{exactly} ground states (or vacua) of \emph{free} scalar QFTs with dispersions rational in $\kk^2$,
\begin{equation}\label{eq_rational_dispersion}
    \freq(\kk) = {P(\kk^2)}/{Q(\kk^2)}
    \ ,
\end{equation}
with $P, Q$ polynomials.
Clearly, this is only a subset of all possible free QFTs;
among the excluded models are all relativistic QFTs, which have
\begin{equation}\label{eq_kg_dispersion}
    \freq(\kk) = \sqrt{\mass^2+\kk^2}
    \ ,
\end{equation}
$\mass$ being the field's mass.
We will return to this point and study the faithfulness of various G\CTNS{} approximations of a free relativistic scalar field in \cref{sec_kg}.
Since valid G\CTNS{} kernels $\kernel$ have no poles at real $\kk$, $\pham$ is still quasi-local;
it becomes local if $\kernel$ is a polynomial ($Q$ constant), in which case
$\ctns$ is a local \emph{Rokhsar-Kivelson} wavefunctional, encoding the partition function of a $\spdim$-dimensional statistical field theory \cite{Rokhsar1988SuperconductivityQuantumHard,Fradkin2006EntanglementEntropy2d,Boudreault2022EntanglementSeparabilityContinuum}.
A physically relevant example of local vacua is provided by free \emph{Lifshitz} field theories \cite{Ardonne2004TopologicalOrderConformal,Alexandre2011LifshitzTypeQuantum} with even dynamical exponent $z$, and certain massive deformations thereof \cite{Boudreault2022EntanglementSeparabilityContinuum}.
Lifshitz models with $z\neq1$ describe non-relativistic scale-invariant systems exhibiting anisotropic scaling between space and time, also known as Lifshitz scaling,
\begin{equation*}
    t \to \lambda^z t
    \ , \quad
    \xxx \to \lambda \xxx
    \ ,
\end{equation*}
with $\lambda > 0$, implying
\begin{equation}\label{eq_lifshitz_dispersion}
    \freq(\kk) \propto \abs{\kk}^z
    \ .
\end{equation}
For a free Lifshitz scalar with positive even $z$,
this gives a local vacuum wavefunctional
represented by a G\CTNS{} with $\cbd=0$ and $(z/2)$-order derivatives or, by \cref{eq_higher_order_count}, $\cbd=z/2-1$ and first-order derivatives.

The rational dispersion \cref{eq_rational_dispersion} always approaches an even power of $\kk$ at large momentum, hence G\CTNS{} look like even-$z$ Lifshitz ground states at short-distance scales.
In the language of the Wilsonian (momentum-shell) renormalization group (RG),
the UV fixed point of an $\dvord$th derivative G\CTNS{} with $\cbd$ virtual fields is a free Lifshitz vacuum with
\begin{equation}\label{eq_rg_fixed_point}
    z = 2\deg P - 2\deg Q \leq \ctnsz
    \ .
\end{equation}
Indeed, when acting on the quadratic wavefunctional $\ctns$, RG flow just rescales the couplings $\ctnspar$ in $\kernel$ according to their engineering dimension; therefore,
its Gaussian UV fixed point is the leading homogeneous piece of $\kernel$ that dominates at large momenta \cite{Wilson1974RenormalizationGroup}.
\Cref{eq_rg_fixed_point} is to be contrasted with relativistic QFTs, where Lorentz invariance implies $z=1$ and the UV fixed point is expected to be a conformal field theory (CFT) \cite{Nakayama2015ScaleInvarianceVs}.

\subsection{Entanglement scaling}\label{sec_entanglement}
A defining property of ordinary (discrete) TNS is that the entanglement they can accommodate is bounded by an increasing function of the bond dimension $\cbd$.
E.g.,
\begin{math}
    \ent(\subsys)\leq\tnarea\log{\cbd}
\end{math}
for the entanglement entropy (EE) of a subsystem $\subsys$, where $\tnarea$ here denotes the area of the boundary in the TN geometry, \idest{}, the minimum number of virtual bonds that have to be cut in order to isolate $\subsys$ (\eg{}, $\tnarea=2$ for a bulk interval of an MPS)
\cite{Cirac2021MatrixProductStates}.
Whether CTNS obey analogous bounds with respect to the number of auxiliary fields is an open question \cite{Tilloy2019ContinuousTensorNetwork}.
The difficulty in answering it comes in part from the fact that
EE is known to be UV divergent in QFT, where every finite patch of space hosts an infinite number of degrees of freedom \cite{Holzhey1994GeometricRenormalizedEntropy,Witten2018ApsMedalExceptional}.
Consequently, \emph{universal} (\idest{}, regularization-independent) entanglement quantifiers must be identified.
Given a local QFT, regularized by means of a short-distance cutoff $\uv$, one expects the following $\uv\to0$ expansion for the vacuum EE of a $\spdim$-dimensional subregion \cite{Casini2009EntanglementEntropyFree}:
\begin{equation}
    \label{eq_entropy_scaling}
    \ent
    =
    g_{\spdim-1}(\subsys^)\,\uv^{-(\spdim-1)} \:+\: \dots \:+\: g_1(\subsys^)\,\uv^{-1}
    - g_0(\subsys^)\,\log (\uv)+ \ent_0(\subsys)
    \,,
\end{equation}
where $\ent_0(\subsys)$ is finite and $g_i$ are local functions of the entangling surface $\subsys^$, which are homogeneous of degree $i$.
The leading power-law divergences are unphysical (\idest{}, regularization-dependent), but the coefficient $g_0$ of the logarithmic divergence (if present) is expected to be regularization-independent and tied to the UV universality class of the QFT. The finite part $\ent_0$ also includes universal contributions which, however, are less straightforward to isolate \cite{Casini2009EntanglementEntropyFree}.

As we just discussed, the UV fixed points of G\CTNS{} are free bosonic Lifshitz vacua with $z\leq\ctnsz$.
Their EE is expected to increase with $z$, as can be argued on the lattice based on the longer-range interactions arising from the discretization of higher-order derivatives \cite{He2017EntanglementEntropyLifshitz}.
In fact, CMERA predictions \cite{Gentle2018LifshitzEntanglementEntropy} and numerical observations \cite{MohammadiMozaffar2017EntanglementLifshitzType} suggest a crossover from an area-law EE scaling at finite $z$, compatible with \cref{eq_entropy_scaling}, to a volume law for $z\to\infty$.
Whether this EE increase only inflates the leading divergence or also reflects on universal quantities (and if so, how)
is not obvious in general.
In one space dimension, this question has been recently answered exploiting the Rokhsar-Kivelson structure of Lifshitz vacua:
for a bulk interval
and $z>1$ \cite{Basak2024MasslessLifshitzField},
\begin{equation}\label{eq_lifshitz_g0}
    g_0 = z-1
    \ ,
\end{equation}
which grows with $z$ as anticipated.\footnote{\label{note_uv_ir_mixing}%
    Although the bulk interval EE is IR-divergent for a massless Lifshitz scalar,
    \begin{math}
        g_0=-{\odv{\ent}/{\log\uv}_{\uv=0}}
    \end{math}
    is independent of the system size $\ir$ and,
    for $z=2$, $g_0$ was also shown to be stable against massive deformations \cite{Boudreault2022EntanglementSeparabilityContinuum}.
    Both features
    follow
    the general expectation that long-distance features of a QFT should not affect its UV behavior and vice versa (no UV/IR mixing),
    and
    mirror the 1+1$\spdim$ relativistic case: Klein-Gordon has the same EE scaling as a conformal boson and finite-size corrections
    are independent of $\uv$ \cite{Calabrese2004EntanglementEntropyQuantum}.
}
This result implies that the EE of 1$\spdim$ G\CTNS{}\footnote{Still equipped with infinite-dimensional virtual spaces, \idest{}, not a CMPS.} has a logarithmic divergence whose coefficient is bounded by a linearly increasing function of $\cbd$.
A numerical verification will be shown in \cref{sec_kg_entanglement}.
In higher dimensions the landscape is more complex:
in 2$\spdim$ exact analytical results for $z=2$ show that EE has no logarithmic UV divergence for cuts across smooth entangling surfaces \cite{Hsu2009UniversalEntanglementEntropy,Zhou2016EntanglementEntropyMutual} but
corner singularities still produce universal log terms \cite{Bueno2016BoundsCornerEntanglement,Boudreault2022EntanglementSeparabilityContinuum}.
Numerical investigations \cite{MohammadiMozaffar2017EntanglementLifshitzType} indicate the same qualitative picture holds for $z>2$;
however, to our knowledge, monotonicity in $z$ of the log-coefficient has not been established.

\section{Approximating ground states}\label{sec_gs}

We now turn to the problem of finding an approximate \CTNS{} representation of the ground state of a given QFT.
A well-established practice for TNS is to treat some \CTNS{} submanifold as a variational class over which to minimize the energy, \eg{} via gradient descent or imaginary-time evolution \cite{Schollwoeck2011DensityMatrixRenormalization,Banuls2023TensorNetworkAlgorithms}.
Variational optimization in the continuum has been successfully carried out with CMPS \cite{Verstraete2010ContinuousMatrixProduct,Tuybens2022VariationalOptimizationContinuous}, RCMPS \cite{Tilloy2022StudyQuantumSinh,Tiwana2025RelativisticContinuousMatrix,Tiwana2025MultiFieldRelativistic}, as well as G\CTNS{} \cite{Karanikolaou2021GaussianContinuousTensor}.
In what follows, we revisit this problem in the context of free theories and G\CTNS{}.
This will allow us to highlight analytically some properties of the latter.

\subsection{Rational approximants}\label{sec_gs_rational}

Minimizing the expectation value of the free Hamiltonian
\begin{equation}\label{eq_free_hamiltonian}
    \ham
    = \xint \:\hdens(\xxx)
    = \frac12\kint \brackets[\Big]{\ppop*\ppop + \freq^2(\kk) \ffop*\ffop}
    \ ,
\end{equation}
with ground state
\begin{equation*}
    \vac(\ff) = \exp*{-\frac{1}{2}\kint\, \freq(\kk)\ff*\ff}
    \ ,
\end{equation*}
over the class of G\CTNS{} with bond dimension $\cbd$ corresponds by \cref{eq_gctns_correlators} to minimizing
\begin{equation}\label{eq_variational_energy}
    \edens(\kernel) - \edens_0 =
    \vev{\hdens}_{{\textstyle\mathstrut}\ctns} -
    \vev{\hdens}_{{\textstyle\mathstrut}\vac}
    =
    \frac{1}{4}\kint \frac{\brackets{\kernel(\kk)-\freq(\kk)}^2}{\kernel(\kk)}
\end{equation}
over the set of rational kernels $\kernel$ of degree $\kerdeg$ in $\kk^2$.
Alternatively, maximizing
the fidelity
\begin{math}
    F = \abs{\braket{\ctns}{\vac}}^2
\end{math}
corresponds, introducing a finite-volume regulator $V$,
to minimizing
\begin{equation}\label{eq_fidelity}
    f=-\frac1V \log F = \kint\log\frac
    {{\freq(\kk)+\kernel(\kk)}}{2\sqrt{\freq(\kk)\kernel(\kk)}}
    \ ,
\end{equation}
in which case $\freq$ can also be an effective dispersion obtained using a Gaussian variational ansatz for an interacting QFT
(Hartree approximation,
shifting $\ff$ so that $\vev{\ffop}=0$) \cite{Stevenson1985GaussianEffectivePotential,Polley1988VariationalCalculationsQuantum}.

Unsurprisingly, both cost functions essentially measure the functional distance between the kernels $\freq$ and $\kernel$.
The problem of finding a G\CTNS{} approximation of $\VAC$ with $\cbd$ virtual fields thus reduces to that of identifying a rational approximant $\kernel$ of $\freq$ of degree $\kerdeg$ in $\kk^2$ defining a valid Gaussian kernel --- namely positive and finite at all real $\kk$, possibly up to a zero at $\kk=0$.
We then know that \cref{eq_gctns_kernel} can be solved for $\ctnsmat$.
If a linear partial fraction or continued fraction representation of $\kernel$ is available, this amounts to reading off the G\CTNS{} coefficients from \cref{eq_rational_simple_poles} or \cref{eq_tridiagonal_cf},
while a general solution is provided by \cref{eq_canonical_jordan}.

A simple example is the approximation of the $\mass>0$ Klein-Gordon vacuum in ref.~\cite{Shachar2022ApproximatingRelativisticQuantum}, defined
by the continued fraction \cref{eq_tridiagonal_cf} with
\begin{equation}\label{eq_pade_cf}
    \{\tridia\}\fvlims = \{\mass, 2\mass \kk^2, 2\mass, 2\mass \kk^2, 2\mass, \ldots\}
    \ , \qquad
    \trioff=\kk^2
    \ ,
\end{equation}
which is generated by recursion from the identity
\begin{equation*}
    \sqrt{\mass^2+\kk^2} \:=\: \mass + \frac{\kk^2}{\mass + \sqrt{\mass^2+\kk^2}}
    \ ,
\end{equation*}
followed by an equivalence transformation
\begin{equation*}
    \mass+\cfrac{\kk^2}{
        2\mass + \cfrac{\kk^2}{
            2\mass + \cfrac{\kk^2}{
                2\mass + \ddots}
        }
    }
    \:=\:
    \mass+\cfrac{\kk^4}{
        2\mass\kk^2 + \cfrac{\kk^4}{
            2\mass + \cfrac{\kk^4}{
                2\mass\kk^2 + \ddots}
        }
    }
    \;\;,
\end{equation*}
to render $\trioff$ polynomial in $\kk^2$,
ensuring that $\ctnsfunc$ is local and thus defines a valid G\CTNS{}.

\begingroup\let\vindmax\evonum

\subsection{Imaginary-time evolution}\label{sec_gs_evo}
The ground state $\vac$ of a gapped Hamiltonian $\ham$ can also be obtained via imaginary-time evolution:
for any state $\evoinit$ having non-vanishing overlap with $\vac$,
\begin{equation*}
    e^{-\evoT \ham} \evoinit> \sim
    \braket{\vac}{\evoinit}
    e^{-\evoT E_0} \ket{\vac}
\end{equation*}
when $\evoT\to\infty$.
Disregarding normalization issues
and
splitting the imaginary-time interval $[-\evoT, 0]$ into $\evonum$ short segments,
\begin{align}\label{eq_evo_trotter}
    \renewcommand{\evoeps}{(\evoT/\evonum)}
    \ket{\vac}
    = \lim_{\evoT\to\infty}\lim_{\evonum\to\infty}
    \int\!\DD\ffv_0\cdots\DD\ffv[-1]
    \op{\ffv[0]} \evostep
    \op{\ffv_1} \cdots \evostep
    \ket{\ffv[-1]}\!\braket{\ffv[-1]}{\evoinit}
    \ .
\end{align}
Comparing with \cref{eq_ctns}, one is tempted to identify the path integral above with a \CTNS{} with $\cbd=\evonum$.
However, even for local Hamiltonians, the integrand is not the exponential of a local functional.
The standard solution is to expand
to first order in the Trotter step, $\evoeps=\evoT/\evonum\ll1$.
Assuming
\begin{math}
    \ham
        = {K(\ppop)} + {V(\ffop)}
\end{math},
\begin{equation*}
    \mel{\ffv<}{\evostep}{\ffv}
    \approx\!
    \int\!\DD\pp
    \braket{\ffv<}{\pp}\!\mel{\pp}{
        e^{-\evoeps K(\ppop)}
        e^{-\evoeps V(\ffop)}
    }{\ffv}
    =\!
    \let\ophat\relax
    \int\!\DD\pp \: e^{
            i\!\xint\, \pp[\ffv<-\ffv] - \evoeps \ham(\pp,\ffv)
        }
    \ .
\end{equation*}
For the usual case of
\begin{math}
    K \mathbin= \int\pp^2/2
\end{math},
the functional integral over $\pp$ is just the value of the integrand at its stationary point.
Plugging into \cref{eq_evo_trotter} and
formally setting
\begin{math}
    \evoinit = 1
\end{math},
so that dependence on the initial state drops out,
the path integral takes the \CTNS{} form \cref{eq_ctns}, with
\begin{math}
    \cbd=\evonum
\end{math}
and
\begingroup\let\ophat\relax%
\begin{equation}\label{eq_evo_ctns}
    \let\ophat\relax
    \ctnsfunc(\fff)
    =
    \evoeps \sum\vvlims \brackets[\Bigg]{
        \frac{1}{2} \xint
        \brackets[\Big]{\frac{\ffv<-\ffv}{\evoeps}}^{\mathrlap{2}}
        + V(\ffv)
    }
    =
    \evoeps \sum\vvlims \ham \parentheses[\Big]{
        \pp\compeq\frac{\ffv<-\ffv}{\evoeps},
        \ff\compeq\ffv
    }
    \ .
\end{equation}
Assuming $\ham$ satisfies the usual properties (locality, translation and rotation invariance) that we required of a good \CTNS{}, $\ctnsfunc$ above defines a family $\evovac$ of \CTNS{} approximations of $\vac$ with the same spatial-derivative order as $\ham$.
In the $\evoT,\evonum\to\infty$ limit, the usual path-integral representation of the QFT vacuum wavefunctional \cite{Casini2009EntanglementEntropyFree} is recovered using
\begin{math}
    \evoeps\sum\vvlims\to\int_{-\evoT}^0\dd{\itime}
\end{math}
and
\begin{math}
    [\ffv<-\ffv]/\evoeps\to\dffdit({-\vvind\evoeps})
\end{math}:
\begin{equation*}
    \lim_{\evoT\to\infty}\lim_{\evonum\to\infty} \vac_{\evonum,\evoT/\evonum}(\ffv[0])
    =
    \int\limits_{\mathclap{\varphi(\itime=0)
            =
            \ffv[0]}} \DD\varphi\:
    e^{-\int_{-\infty}^{0} \dd{\itime} \ham(\dffdit(\itime),\varphi(\itime))}
    =
    \int\limits_{\hspace{-2em}\mathrlap{\substack{\varphi(\itime=0) = \ffv[0]\\\itime\in(-\infty,0]}}}\!
    \DD\varphi\:
    e^{-S_E(\varphi)}
    \ ,
\end{equation*}
where $S_E$ is the Euclidean action and $\varphi$ is now a Euclidean field, \idest{} a function of both space and imaginary time.
\endgroup

\CTNS{} virtual fields may be interpreted as sampling physical field configurations at past imaginary times, and $\ctnsfunc$ only intertwines fields from consecutive time steps.
Indeed, $\evovac$ is obtained simply applying the (Trotter expanded) imaginary-time propagator $\evostep$ to $\vac_{\evonum-1,\evoeps}$.
The decomposition
\begin{math}
    \evostep = \trotter[2]{\HK,\HV}
\end{math}
used above is clearly not unique;
for instance, using
\begin{math}
    \evostep \approx \trotter{(1-\trottercoeff)\HV,\HK,\trottercoeff\HV}
\end{math}
one gets
\begin{equation}\label{eq_evo_step_ctns}
    \vac_{\evonum,\evoeps}(\ff) =
    \int\DD\fftilde\:
    e^{
            -\frac{1}{2\evoeps}\xint(\ff-\fftilde)^2
            - \evoeps(1-\trottercoeff) V(\ff) - \evoeps\trottercoeff V(\fftilde)
        }\:
    \vac_{\evonum-1,\evoeps}(\fftilde)
    \ , \quad
    \vac_{0,\evoeps} = \evoinit=1
    \ ;
\end{equation}
$\trottercoeff=1/2$ gives the standard second-order Suzuki-Trotter decomposition.
All decompositions lead to the same $\evoeps\to0$ limit, but they differ at finite $\evoeps$ (see \eg{} \cref{tab_asymptotics} on \cpageref{tab_asymptotics}).

For the free Hamiltonian \cref{eq_free_hamiltonian},
$\ctnsfunc$ in \cref{eq_evo_ctns} defines a tridiagonal Gaussian \CTNS{}.
The same holds for its generalization \cref{eq_evo_step_ctns}, as is also seen directly from the resulting kernel $\evoker$: let
\begin{equation}\label{eq_trotter_coeffs}
    \evodia_0=1+(1-\trottercoeff)\evoeps^2\freq^2
    , \quad
    \evodia_1 = 1+\trottercoeff\evoeps^2\freq^2
    , \quad
    \evodia
    =(\evodia_0+\evodia_1)/2
    =1+\evoeps^2\freq^2/2
    \ ;
\end{equation}
carrying out the Gaussian path integral, \cref{eq_evo_step_ctns} amounts to
\begin{equation}\label{eq_evo_step_kernel}
    \evoeps\evoker =
    \evodia_0 - \brackets{\evodia_1+\evoeps\kernel_{\evonum-1,\evoeps}}\inv
    \ , \qquad
    \kernel_{0,\evoeps}=0
    \ .
\end{equation}
By recursion, this generates a continued fraction \cref{eq_tridiagonal_cf} with uniform time-bulk coefficients:\footnote{%
    Here $\trioff$ is imaginary because imaginary-time evolution couples time slices through a real term,
    the cross term of the discretized kinetic energy $(\ffv<-\ffv)^2/2\evoeps$,
    whereas \cref{eq_gctns_tridiagonal} factors an $i$ out of the off-diagonal couplings, so that nested fractions in \cref{eq_tridiagonal_cf} have a plus sign.
}
\begin{equation}\label{eq_evo_W}
    \{\evoeps\tridia\}\fvlims = \{\evodia_0,2\evodia,2\evodia,\ldots,\evodia_1\}
    \ , \qquad
    \evoeps\trioff=i
    \ ,\qquad
    (\evonum>0)
    \ .
\end{equation}
The $\evoeps$ prefactors can be dropped by substituting
\begin{math}
    \kernel\to\evoeps\kernel
\end{math}
in \cref{eq_tridiagonal_cf}.
Here we restricted to product formulas of type \cref{eq_evo_step_ctns}.
Higher-order formulas still generate tridiagonal G\CTNS{}, but introduce multiple virtual fields per time step and generally result in continued fractions with period greater than one, due to their non-uniform stage coefficients.
Convergence of the G\CTNS{} approximation is verified, for first- and second-order formulas, by recognizing in \cref{eq_evo_W} the continued fraction expansion of a square root:\footnote{%
    To simplify the derivation, we reversed the order of the limits with respect to \cref{eq_evo_trotter}
    and assumed the last denominator $\tridia[-1]$ is irrelevant in the $\evonum\to\infty$ limit;
    both assumptions are justified in \cref{sec_cf_convergence}.
}
\begin{equation}\label{eq_evo_convergence}
    \lim\limits_{\evoeps\to0}
    \lim\limits_{\evonum\to\infty}\evoker
    =
    \lim\limits_{\evoeps\to0}
    \evoeps\inv \brackets*{\sqrt{\evodia^2-1}+ \evodia_0 - \evodia}
    =
    \freq
    \lim\limits_{\evoeps\to0}
    \brackets*{\sqrt{1 + \frac{\evoeps^2\freq^2}{4}} + C_1 \evoeps\freq}
    =\freq
    \ ,
\end{equation}
where $C_1=(\evodia_0-\evodia)/\evoeps^2\freq^2$, \idest{}, $C_1=\mp1/2$ for first-order decompositions and $C_1=0$ for the second-order one, consistent with an $\order{\evoeps^2}$ Trotter error on the final state.

\paragraph{Compressing the G\CTNS{}.}
There is an important (yet expected) lesson to be learnt: as implemented above, without some compression scheme, imaginary-time evolution will generally not give the optimal \CTNS{} approximation of the vacuum with a given bond dimension, because in \cref{eq_evo_ctns} $\cbd$ is just the number of Trotter steps, which has no physical meaning.
Consider for instance a free QFT with
\begin{math}
    \freq(\kk) = \mass^2 + \kk^2
\end{math}, \idest{}, a massive deformation of a $z\mathbin=2$ Lifshitz field.\footnote{%
    Lifshitz scaling implies the absence of a gap, preventing the convergence of imaginary-time evolution, however such massive deformation preserves the locality of $\VAC$ and has been considered in the literature \cite{Boudreault2022EntanglementSeparabilityContinuum}.
}
Although its vacuum admits an exact $\cbd=0$ first-derivative G\CTNS{} representation, \cref{eq_evo_W}
converges to the true ground state only for $\evoT,\cbd/\evoT\to\infty$.
Moreover, since $\freq^2$ enters the G\CTNS{} coefficients \cref{eq_trotter_coeffs}, the derivative order is twice as large as needed in this case.

Standard TN implementations of imaginary-time evolution, such as time-evolving block decimation \cite{Vidal2004EfficientSimulationOne} or the time-dependent variational principle \cite{Haegeman2011TimeDependentVariational,Vanderstraeten2019TangentSpaceMethods,Hackl2020GeometryVariationalMethods}, feature a truncation or projection step to keep the state within some chosen TNS manifold and its bond dimension under control.
Additionally, a well-conditioned optimization should converge to the exact ground state when the latter admits an exact representation within the chosen manifold.
The same holds for \CTNS{}, as we now illustrate with a simple example, using again the $z=2$ massive Lifshitz theory as a toy model.

\begin{figure}[b]
    \includegraphics{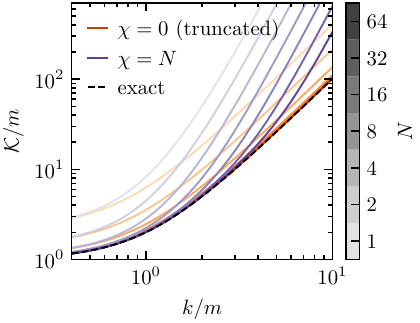}
    \caption{%
        Imaginary-time evolution G\CTNS{} kernels $\kernel$ for a massive deformation of the $z\mathbin=2$ free Lifshitz field theory, \emph{with \cref{eq_evo_truncation_lifshitz} and without \labelcref{eq_evo_W} truncation};
        second-order $\trottercoeff\mathbin=1/2$ decomposition, $\mass\evoT\mathbin=4$.
        The truncated $\kernel$ becomes visually indistinguishable from the exact dispersion $\freq(\protect\kk) = \mass^2 + \protect\kk^2$, depicted by a black dashed line, already at $\evonum\ge16$.
    }
    \label{fig_lifshitz}
\end{figure}

To stay within a given \CTNS{} manifold during the evolution means that the state $\evovac$ keeps the form \cref{eq_ctns} with a fixed functional $\ctnsfunc$ and, after each time step, only its parameters $\ctnspar=\ctnspar_\evonum$ change.
In the Gaussian case, we can devise an update rule for evolving a G\CTNS{} under a quadratic Hamiltonian by complementing each Trotter step \cref{eq_evo_step_kernel} with a rational approximation $\truncproj$ of fixed degree:
\begin{equation}\label{eq_evo_truncation}
    \evoeps\evoker
    =
    \truncproj\brackets*{
        \evodia_0 - \frac{1}{\evodia_1+\evoeps\kernel_{\evonum-1,\evoeps}}
    }
    \ .
\end{equation}
Write the kernel in reduced form as $\kernel\mathbin=P/Q$.
While $\deg Q\mathbin<\cbd$, the projector $\truncproj$ acts trivially and each recursion increases $\deg Q$ by one, thus adding a new virtual field. Once $\deg Q\mathbin=\cbd$, $\truncproj$ keeps the state within the chosen G\CTNS{} manifold.

In the simplest example of $\cbd=0$,
$\evovac$ is parameterized by $\ctnspar_\evonum\mathbin=(\trunca,\truncb)$ via
\begin{equation}\label{eq_evo_truncation_lifshitz}
    \evoker(\kk) = \trunca + \truncb\kk^2
    \ .
\end{equation}
Taking $\truncproj$ to be the Taylor expansion to order $\kk^2$ around $\kk=0$ yields
\begin{equation*}
    \evoeps\trunca> + \evoeps\truncb>\kk^2 =
    \evodia_0(0)-\frac{1}{\evodia_1(0)+\evoeps\trunca}
    + \frac12
    \brackets*{
        \evodia_0''(0)
        + \frac
        {\evodia_1''(0) + 2\evoeps\truncb}
        {(\evodia_1(0) + \evoeps\trunca)^2}
    } \kk^2
    \ .
\end{equation*}
Matching coefficients order-by-order, we get recurrence relations for $\trunca$ and $\truncb$:
\begin{equation}\label{eq_evo_truncation_recurrence}
    \brackets{\evoeps\trunca>-\evodia_0(0)}
    \brackets{\evoeps\trunca+\evodia_1(0)} = -1
    \ ,\quad
    \frac
    {\truncb>-\evoeps(1-\trottercoeff)\gamma}
    {\truncb+\evoeps\trottercoeff\gamma}
    =-
    \frac
    {\evoeps\trunca>-\evodia_0(0)}
    {\evoeps\trunca+\evodia_1(0)}
    \ ,
\end{equation}
with coefficients determined by the physical dispersion $\freq$ through $\evodia_{i}(0)$ and $\gamma=\freq(0)\,\freq''(0)$,
and initial conditions set by the initial state $\evoinit$.
Setting $\evoinit=1$ as usual, $\trunca!=\truncb!=0$.
Upon solving these recurrence relations, one gets an update rule for the $\ctnspar_\evonum$ G\CTNS{} parameters that projects imaginary-time evolution on the $\cbd=0$ manifold.
\Cref{fig_lifshitz} shows that, for a $z=2$ Lifshitz theory, $\kernel$ converges to $\freq$ for $\evonum\to\infty$ even at $\cbd=0$, as expected.
Coincidentally, since in this case the exact dispersion is completely characterized by $\freq(0)$ and $\freq''(0)$, the truncation actually accelerates convergence at high momenta.

\endgroup

\section{Free relativistic boson}\label{sec_kg}

The Klein-Gordon theory of a free scalar field is the simplest example of relativistic QFT.
It provides the ideal playground for studying the limitations of G\CTNS{} approximations of relativistic vacua.
These stem from the clash between the conformal UV of relativistic QFT and the even-$z$ Lifshitz UV \cref{eq_rg_fixed_point} of G\CTNS{}: an infinite number of high-frequency modes are misrepresented by any finite-$\cbd$ G\CTNS{} for Klein-Gordon, because $\kernel$ rational in $\kk^2$ can never reproduce $\freq \sim \kk$ asymptotically.
Accordingly, a UV cutoff must be introduced to stabilize variational optimization \cite{Tilloy2021VariationalMethodRelativistic,Tilloy2021RelativisticContinuousMatrix,Stojevic2015ConformalDataFinite,Haegeman2010ApplyingVariationalPrinciple}.
Without it, the energy \cref{eq_variational_energy} and fidelity \cref{eq_fidelity} cost functions are ill-defined and,
in fact,
G\CTNS{} do not even live in the same Fock space as the free relativistic boson because the latter has an infinite density of excitations with respect to any G\CTNS{}, violating the Shale criterion \cite{Tilloy2021RelativisticContinuousMatrix,Shale1962LinearSymmetriesFree}:
\begin{equation}\label{eq_shale}
    \int\!\!\dd[\spdim]{\kkk}\: \abs{\beta(\kk)}^2
    =
    \int\!\!\dd[\spdim]{\kkk}\: \frac{\brackets{\kernel(\kk)-\freq(\kk)}^2}{4\,\freq(\kk)\kernel(\kk)} < \infty
    \ ,
\end{equation}
where the squared Bogoliubov coefficient $\abs{\beta}^2$ measures the momentum-space density of G\CTNS{} excitations in the relativistic vacuum, and vice versa.

Ultimately, similarly to a lattice regularization, G\CTNS{} can thus accurately capture the low-energy, long-distance properties of the field theory, but introduce a UV scale $\pepsuv$.
Fortunately, the latter grows \emph{exponentially} with the bond dimension!
We make this more precise below ---
but first, what if one insists on realizing the relativistic dispersion through the parent Hamiltonian $\pham$ \cref{eq_parent}?

\subsection{Non-local field bases}\label{sec_kg_nonlocal}

Dropping the canonical normalization of the conjugate momentum in the parent Hamiltonian \cref{eq_parent} in favor of
\begin{math}
    \freq(\kk) = \sqrt{\mass^2+\kk^2}
\end{math},
$\pham$ is still \emph{not} the Klein-Gordon Hamiltonian
\begin{equation*}
    \ham[KG] = \frac{1}{2}\xint\:\brackets[\Big]{
        \ppnew^2 + (\nabla\ffnew)^2 + \mass^2\ffnew^2
    }
    \ ;
\end{equation*}
it is only equivalent to it through the \emph{non-local} canonical transformation
\begin{equation}\label{eq_kg_canonical_transformation}
    \ffnew(\kkk) = \sqrt{\frac{\kernel(\kk)}{\freq(\kk)}}\,\ffold(\kkk)
    \ , \qquad
    \ppnew(\kkk) = \sqrt{\frac{\freq(\kk)}{\kernel(\kk)}}\,\ppold(\kkk)
    \ .
\end{equation}
Substituting in $\ctns$ in \cref{eq_gctns_contracted} replaces $\kernel$ with $\freq$.
Hence,
a G\CTNS{} can encode Klein-Gordon's vacuum, just in a non-standard field basis.
However, no valid choice of $\kernel$ can make the transformation \cref{eq_kg_canonical_transformation} local, since $\kernel$ is rational and $\freq$ is not,
nor is it unitarily implementable, since by \cref{eq_shale} $\ffnew$ and $\ffold$ do not act on the same Fock space.
The Klein-Gordon field $\ffnew(\xxx)$ is then obtained from $\ffold(\xxx)$ via convolution with a quasi-local kernel
and, \eg{},
expectation values of field monomials $\vev{\ffnew^n(\xxx)}$ and correlators $\vev{\ffnew(\xxx_1)\cdots\ffnew(\xxx_n)}$ translate
into integrals of $n$-point functions for $\ffold(\xxx)$, which can be evaluated directly from the \CTNS{}.\footnote{%
    For the free theory, one could define the \CTNS{} in $\kkk$-space, where \cref{eq_kg_canonical_transformation} is local,
    but, ultimately, the goal is to study interacting theories, where the Hamiltonian itself becomes non-local in momentum space.
}
This is essentially how the RCMPS ansatz works, at least in its most naïve implementation \cite{Tilloy2021RelativisticContinuousMatrix}.
Its ``computational'' field basis (or the higher-dimensional analogue) is the Fourier transform of the creation-annihilation pair $\fock*(\kkk), \fock(\kkk)$ in the mode expansion of $\ffnew$,
\begin{equation}
    \fock(\xxx) = \kint
    \sqrt{\frac{\freq(\kk)}{2}}\brackets*{\ffnew(\kkk) + \frac{i}{\freq(\kk)}\ppnew(\kkk)}
    \ .
\end{equation}
Setting $\kernel=\ladderscale$ with $\ladderscale$ some arbitrary energy scale, $\fock*(\xxx)$ and $\fock(\xxx)$ are nothing but the non-relativistic creation and annihilation operators for the \CTNS{} computational field:
\begin{equation}
    \ffold(\xxx) = \sqrt{\frac{1}{2\ladderscale}}\brackets[\big]{\fock*(\xxx) + \fock(\xxx)}
    \ , \qquad
    \ppold(\xxx) = i\sqrt{\frac{\ladderscale}{2}}\brackets[\big]{\fock*(\xxx) - \fock(\xxx)}
    \ .
\end{equation}
The advantage of the RCMPS Bogoliubov basis is that it allows one to straightforwardly normal order every operator, which in one dimension is enough to remove virtually all UV divergences \cite{Tilloy2021RelativisticContinuousMatrix}.
In higher dimensions, this is no longer the case and, for interacting theories, a cutoff is needed anyway.
For this reason, here we refrain from any non-local field redefinition, and identify the \CTNS{} field with the physical one.

\subsection{UV scale and bond dimension}\label{sec_kg_dispersion}

\begin{table}[p]
    \def\evohead{\multicolumn{3}{c}{Trotter $\kernel=\evofreq$ \cref{eq_evo_W}}}
    \def\padehead{\multicolumn{2}{c}{Padé $\kernel=\padefreq$ \cref{eq_pade_cf}}}
    \begin{tabular}{*{6}{>{$}c<{$}}}
        \toprule
                                 & \padehead       & \evohead                                                                                                         \\
        \cmidrule(lr){2-3}\cmidrule(l){4-6}
                                 & \cbd\text{-odd} & \cbd\text{-even} & q=1                           & q=0               & q=1/2                                     \\
                                 &                 &                  & \trotter{\HK,\HV}             & \trotter{\HV,\HK} & \trotter{\frac{\HV}{2},\HK,\frac{\HV}{2}} \\
        \midrule
        \pepsuv\text{ (UV)}      & \mass\cbd       & \mass\cbd        & 1/\evoeps                     & 1/\evoeps         & 2/\evoeps                                 \\
        \pepsfreq(\kk\gg\pepsuv) & \kk^2/\mass\cbd & \mass\cbd        & 1/\evoeps                     & \evoeps\kk^2      & \evoeps\kk^2/2                            \\
        \midrule
        \pepsir\text{ (IR)}      & 0               & 0                & \sqrt{1-\evoT^2\mass^2}/\evoT & \cdots            & \cdots                                    \\
        \pepsfreq(\kk\ll\pepsir) & -               & -                & \evoT(\mass^2+\kk^2)          & \cdots            & \cdots                                    \\
        \bottomrule
    \end{tabular}
    \caption{\label{tab_asymptotics}%
        Coefficients, asymptotes, and domain of convergence of Padé and Trotter G\CTNS{} approximants $\pepsfreq$ of Klein-Gordon's dispersion relation $\freq$.
        The approximation's IR and UV scales, $\pepsir$ and $\pepsuv$, are obtained by matching the corresponding asymptote of $\pepsfreq$ to $\freq$, so that
        \begin{math}
            \pepsfreq(\protect\kk)\approx\freq(\protect\kk)
        \end{math}
        for
        \begin{math}
            \pepsir\ll\protect\kk\ll\pepsuv
        \end{math}.
        Suzuki-Trotter decompositions \cref{eq_evo_step_ctns} with $\trottercoeff\in\{0,1/2,1\}$ are all equivalent in the IR.
        The Padé approximant faithfully reproduces $\freq$ in the IR all the way down to $\protect\kk=0$.
    }
\end{table}
\begin{figure}[p]
    \includegraphics{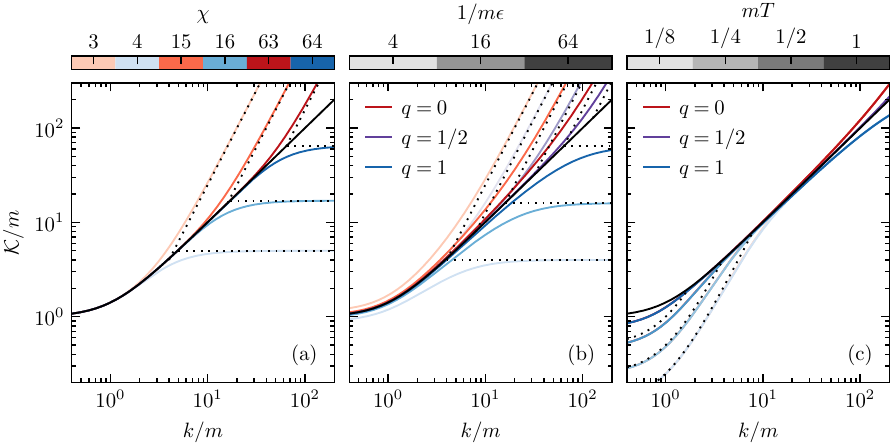}
    \caption{%
        (a) Padé \cref{eq_pade_cf} and (b), (c) Trotter \cref{eq_evo_W} G\CTNS{} approximants $\pepsfreq$ of the Klein-Gordon dispersion relation $\freq$ (black line) for different $\cbd, \evoeps, \evoT$.
        In (b) and (c), different colors denote different decompositions, darker lines correspond to more Trotter steps $\evonum$.
        In (b), the evolution time $\mass\evoT=8$ is fixed and the Trotter step duration
        \begin{math}
            \evoeps=\evoT/\evonum
        \end{math}
        shrinks with
        \begin{math}
            \evonum\in\{32,128,512\}
        \end{math}.
        In (c), $\mass\evoeps=1/256$ is fixed
        and
        \begin{math}
            \evoT=\evoeps\evonum
        \end{math},
        with
        \begin{math}
            \evonum\in\{32,64,128,256\}
        \end{math};
        notice that all decompositions overlap in the IR.
        Trotter G\CTNS{} truncated to $\cbd=10$
        (see main text) are visually indistinguishable from the curves in (b) and (c).
        Dotted lines show the asymptotes in \cref{tab_asymptotics}.
    }
    \subcapanchors{fig_dispersion_pade,fig_dispersion_evo_uv,fig_dispersion_evo_ir}%
    \label{fig_dispersion}
\end{figure}

We now study more systematically various rational approximants of the relativistic dispersion, starting from those found in \cref{sec_gs}.
\Cref{fig_dispersion} shows that they improve for larger $\cbd$ or $\evonum$, as expected; however, convergence is not uniform \cite{Shachar2022ApproximatingRelativisticQuantum}:
errors generally increase with larger momenta and smaller masses --- \idest{}, approaching the CFT regime.
A detailed analysis is found in \cref{sec_cf_convergence}.
Here we focus on the asymptotic behavior, summarized in \cref{tab_asymptotics}, and show how these approximants implicitly define a UV scale.

\paragraph{Padé approximants.}
As we show in \cref{sec_cf_convergence},
the continued fraction kernel \cref{eq_pade_cf}, from now on denoted $\padefreq$, is a Padé approximant of order $[\lceil\cbd/2\rceil/\lfloor\cbd/2\rfloor]$ in $\kk^2$, centered at $\kk=0$.
It thus starts exact at $\kk=0$ \cite{Baker1996PadeApproximants}:
\begin{equation}\label{eq_pade_ir}
    \relerr
    = \frac{\padefreq-\freq}{\freq}
    = \order{(\kk/\mass)^{2(\cbd+1)}}
    \ .
\end{equation}
As the energy increases, deviations appear and $\padefreq\to\freq$ by monotone oscillation as $\cbd\to\infty$, with even-$\cbd$ approximants (blue lines in \cref{fig_dispersion_pade}) converging from below, and odd-$\cbd$ ones (red) from above.
When $\kk/\mass\to\infty$,
\begin{equation}\label{eq_pade_asym}
    \padefreq \sim
    \begin{cases}
        \kk^2/\mass(\cbd+1) & \cbd\text{ odd}  \\
        \mass(\cbd+1)       & \cbd\text{ even}
    \end{cases}
    \ ,
\end{equation}
as can be checked by induction, recalling
\begin{math}
    \padefreq[0] = \mass
\end{math},
\begin{math}
    \padefreq[1] \sim \kk^2/2\mass
\end{math},
and using
\begin{equation*}
    \padefreq[\cbd+2] = \mass + \cfrac{\kk^2}{2m+\cfrac{\kk^2}{\mass+\padefreq}}
    \;\;.
\end{equation*}
Comparing \cref{eq_pade_asym} with the exact dispersion $\freq$ suggests that the Padé approximation breaks down at the energy scale $\pepsuv=\order{\mass\cbd}$, which is confirmed also by \cref{fig_dispersion_pade}.

\paragraph{Trotter approximants.}
Analogous results hold for imaginary-time evolution ground states:
their kernel $\evofreq=\evoker$ \cref{eq_evo_W} approximates $\freq$ over a momentum window which grows with longer evolution times $\evoT$, finer Trotter steps $\evoeps$, and higher-order decompositions.
Asymptotes are again found by induction and matched to $\freq$ to estimate the domain of convergence.
Letting $X=\evoeps\freq$, $Y_{\evonum}=\evoeps\evofreq$, and $\evoinit=1$ as usual, the recurrence \cref{eq_evo_step_kernel} reads
\begin{equation*}
    Y_{\evonum+1}
    =
    \evodia_0 - \brackets{\evodia_1+Y_{\evonum}}\inv
    =
    1+(1-\trottercoeff)X^2 - \brackets{1+\trottercoeff X^2+Y_{\evonum}}\inv
    \ , \qquad
    Y_0 = 0
    \ .
\end{equation*}
Depending on $\trottercoeff$, at least one of $\evodia_0,\evodia_1\ge1$ diverges as $X\to\infty$,
hence,
if $Y_{\evonum} \sim \evodia_0$, the same holds for $Y_{\evonum+1}$,
and since
\begin{math}
    Y_1 = \evodia_0 - 1/\evodia_1 \sim \evodia_0
\end{math},
one concludes that
\begin{math}
    \evofreq \sim \evodia_0/\evoeps
\end{math}
in the UV ($\kk\gg1/\evoeps$, \cref{fig_dispersion_evo_uv})
for all $\evonum>0$.
Matching to $\freq$ gives the UV scale $\pepsuv=\order{1/\evoeps}$.
The UV properties of the G\CTNS{} are thus fully determined by the last Trotter step, namely by $\evoeps$ and $\trottercoeff$, with the second-order decomposition achieving the same UV scale as the first-order ones with half the Trotter steps $\evonum$ (at equal $\evoT$).
Unlike Padé approximants, Trotter approximants are not centered at $\kk=0$; rather, $\evoT$ controls their quality in the infrared (\cref{fig_dispersion_evo_ir}):
\begin{math}
    Y_{\evonum+1}\sim Y_{\evonum}+X^2
\end{math}
for $X,Y_{\evonum}\ll1$, hence
\begin{math}
    Y_{\evonum}\sim \evonum X^2
\end{math}
and thus
\begin{math}
    \evofreq \sim \evoT\freq^2
\end{math}
as $\evoT\freq\to0$.
Once $\evoT\freq\approx1$, the IR expansion breaks down
(see also the comment at the end of \cref{sec_cf_convergence_evo})
and $\evofreq$ crosses over to $\freq$;
only if $\evoT\mass \gg 1$ is the IR cutoff effectively removed.
Note that these asymptotes are general and apply to any quadratic Hamiltonian, such as the massive Lifshitz one of \cref{fig_lifshitz};
their prediction is specialized to Klein-Gordon in \cref{tab_asymptotics}.

We realize the compression $\truncproj$ in \cref{eq_evo_truncation} using the AAA least-squares algorithm for barycentric rational approximation \cite{Nakatsukasa2018AaaAlgorithmRational}.
AAA selects support points from a grid over a prescribed approximation window and returns a rational function of degree at most $(\cbd,\cbd)$, corresponding to a G\CTNS{} with no physical-field gradient term.
We take $\kk\in[0,1/\evoeps]$ as the approximation window: for $\kk\gg1/\evoeps$, the untruncated kernel reduces to its UV asymptote, which only fixes the leading rational coefficient and provides no further physical information or structure to fit.
To preserve the UV asymptotes in \cref{tab_asymptotics}, for $\trottercoeff\mathbin\in\{0,1/2\}$ we run AAA on the kernel divided by $\zeta=1+\kk^2/\mass^2$, then restore this factor to recover degree $\kerdeg$.
We use \textsc{SciPy}'s implementation of AAA~\cite{Virtanen2020Scipy10Fundamental}, with
\texttt{rtol=1e-9}, \texttt{clean\_up\_tol=1e-7}, and
$4000$ points geometrically spaced in $\zeta$,
allowing up to $\cbd=10$ auxiliary fields (\texttt{max\_terms=11}), although the fit often converges with fewer.
For the approximants in \cref{fig_dispersion_evo_uv,fig_dispersion_evo_ir}, the truncated kernels agree with the untruncated ones at equal $\evonum$ within a relative error of at most $\order{10^{-6}}$ over the fitted window, and all their poles lie at $\kk^2<-\mass^2$.

\begin{figure}
    \includegraphics{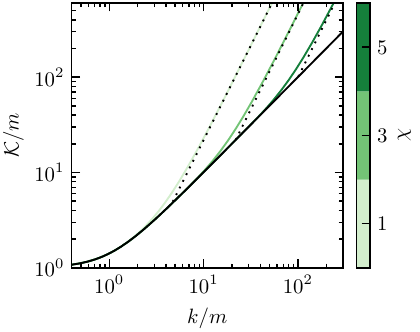}
    \caption{%
        Zolotarev G\CTNS{} approximants \cref{eq_zolo_dispersions} of Klein-Gordon's $\freq$ (black line).
        Same momentum range as in \cref{fig_dispersion}, but $\cbd\leq5$.
        For each $\cbd$, we choose $\pepsuv$ (approximation window) so that the uniform relative error is $\unirelerr=10^{-5}$.
        $\pepsuv$ grows by a factor $e^{\pi^2/\log(4/\unirelerr)}\approx2.2$ per auxiliary field, rather than linearly as in \cref{fig_dispersion_pade}.
        Dotted lines mark the UV asymptotes $\zolofreq \sim d_0 \protect\kk^2/\mass$ \cref{eq_zolotarev}.
    }
    \label{fig_dispersion_zolotarev}
\end{figure}

\paragraph{Zolotarev's optimal approximant.}
We saw that $\cbd=\order{\pepsuv/\mass}$ for Padé approximants.
More precisely, since $\relerr(\kk)$ in \cref{eq_cf_err} grows monotonically with $\kk$, a small uniform error $\unirelerr$ up to $\kk=\pepsuv\gg\mass$ requires
\begin{equation}\label{eq_pade_cbd}
    \cbd \approx \log(2/\unirelerr)(\pepsuv/2\mass)
\end{equation}
by \cref{eq_cf_err_expansions}.
The truncated Trotter approximants found above, however, suggest that this $\cbd$ may not be necessary.
It is then natural to ask what is the \emph{intrinsic} cutoff of a G\CTNS{} with bond dimension $\cbd$;
equivalently,
up to which $\kk^2$ can the ``best'' rational approximation of $\sqrt{\mass^2+\kk^2}$ of degree $\kerdeg$ achieve some target accuracy?
The answer depends on how the approximation error is distributed across energy scales.
\Cref{eq_pade_ir,eq_pade_cbd} show that matching $\freq$ at $\kk=0$ to the highest possible order imposes a cutoff linear in $\cbd$.
Minimizing instead the maximum relative error over $\kk\in[0,\pepsuv]$ leads to a problem solved by Zolotarev in 1877 using Jacobi elliptic functions \cite{Zolotarev1877ApplicationEllipticFunctions}.
We review Zolotarev's construction in \cref{sec_zolotarev}.
It provides an \emph{optimal} approximant $\zolofreq$ \cref{eq_zolo_dispersions} realizable as a G\CTNS{} with $\cbd$ auxiliary fields.
By \cref{eq_zolotarev_error}, at large $\cbd$ and $\pepsuv/\mass$, its uniform relative error $\unirelerr$ decays geometrically in
$\cbd$ at an inverse rate $\propto\log\pepsuv$ (as opposed to $\pepsuv$ for Padé approximants).
Equivalently,
at fixed accuracy, the number of auxiliary fields grows only \emph{logarithmically}
with the cutoff:
\begin{equation}\label{eq_zolotarev_cbd}
    \cbd \approx
    \frac{1}{\pi^2}\log(4/\unirelerr)\log(4\pepsuv/\mass) - 1
    \ .
\end{equation}
The price is a \emph{controlled} error at $\kk=0$: $\zolofreq(0)=(1+\unirelerr)\mass$.
More generally, $\unirelerr$ controls mode by mode the relative errors of the two-point functions in \cref{eq_gctns_correlators}, which in turn determine all higher-order correlators through Wick's theorem.
Zolotarev's approximation is not only favorable in the $\cbd\to\infty$ asymptotics but also at small $\cbd$, as illustrated in \cref{fig_dispersion_zolotarev}.
To facilitate the comparison with \cref{fig_dispersion}
and highlight $\pepsuv\sim\exp\cbd$,
for each $\cbd$ in \cref{fig_dispersion_zolotarev}, we tune $\pepsuv$ to the largest value compatible with $\unirelerr=10^{-5}$.
In practice, one typically proceeds in the opposite direction: increasing $\cbd$ to improve the accuracy at fixed cutoff $\pepsuv$, possibly removing the latter only at the end;
$\zolofreq$ then prescribes how the G\CTNS{} parameters flow with $\pepsuv$, quantitatively implementing a proposal from ref.~\cite{Shachar2022ApproximatingRelativisticQuantum}.

\paragraph{Variational optimization.}
By optimality, for any given UV cutoff, Zolotarev's approximant identifies the best G\CTNS{} \emph{by uniform relative error} in $\kernel$.
In this sense, the logarithmic $\cbd$ in \cref{eq_zolotarev_cbd} is intrinsic to G\CTNS{}.
However, as we now explain, this $\unirelerr$-optimal approximant is not in general obtained by minimizing the energy \cref{eq_variational_energy} or infidelity \cref{eq_fidelity}.
Writing $\kernel=(1+\relerr)\freq$,
\begin{equation}\label{eq_zolotarev_EF_bounds}
    \edens-\edens_0
    = \frac14 \int\limits_{\klims}\!\kint*
    \frac{\relerr^2(\kk)}{1+\relerr(\kk)} \freq(\kk)
    \ , \qquad
    f
    = \int\limits_{\klims}\!\kint*
    \log\cosh\log \sqrt{1+\relerr(\kk)}
    \ ,
\end{equation}
show that, with the regulator in place,
the relative energy excess and infidelity per mode
are also bounded by
\begin{math}
    \unirelerr =\norm{\relerr}_{\infty}
\end{math}
(the converse, of course, is not true).
For example,
evaluating the former on a $\unirelerr$-optimal trial state,
\cref{eq_zolotarev_error} gives
\begin{equation*}
    \relerr_{\edens} =
    \frac{\edens-\edens_0}{\edens_0}
    \leq
    \frac{\unirelerr^2}{2(1-\unirelerr)}
    \approx
    8\exp*{-\frac{2\pi^2(\cbd+1)}{\log(4\pepsuv/\mass)}}
    \ ,
\end{equation*}
implying exponential convergence of the energy with $\cbd$ at fixed cutoff,
and $\cbd\lesssim\log\pepsuv$ at fixed energy accuracy.
These are only upper bounds:
apart from the distinct $L^2$ and $L^\infty$ norms, energy and fidelity optimization weight each momentum shell by a phase-space factor proportional to $\kk^{\spdim-1}\freq(\kk)$ or $\kk^{\spdim-1}$, whereas $\unirelerr$ weights all momenta uniformly.
A target accuracy in either metric might thus be achieved with fewer auxiliary fields than required by a uniformly accurate G\CTNS{}, by fitting momenta close to the cutoff and concentrating the error in the IR, where the phase-space factor is small.
Conversely, to reach a prescribed uniform relative accuracy in $\kernel$, energy and fidelity optimization will generally require more auxiliary fields than a $\unirelerr$-optimal G\CTNS{}, and the scaling with the cutoff may change.

\begingroup
\let\ophat\relax
\paragraph{Interacting theories.}
Finally, although the logarithmic scaling \cref{eq_zolotarev_cbd} concerns a free field,
we expect a similar scaling of $\cbd$ with $\pepsuv$ in Gaussian \CTNS{} approximations of interacting relativistic scalar fields whose short-distance physics remains free.
The UV mismatch in this case must be phrased in terms of equal-time correlators computed in the G\CTNS{} and the target QFT.
For super-renormalizable QFTs, the free-field decay
\begin{math}
    \vev{\ffop\ffop}\sim1/\kk
\end{math}
persists asymptotically while higher-order connected correlators are suppressed at high energy; the same holds up to logarithmic corrections in asymptotically free QFTs.
Matching the two-point function to uniform relative accuracy
with a G\CTNS{} therefore gives $\cbd\sim\log\pepsuv$ in the former case and suggests the same leading scaling in the latter.
This is only a \emph{UV scaling} result: capturing the interacting low-energy physics may still require a cutoff-independent number of additional auxiliary fields, even though we expect a few of them to saturate the best Gaussian approximation of the interacting vacuum, as observed in the non-relativistic Lieb-Liniger model \cite{Karanikolaou2021GaussianContinuousTensor}.
The primary limitation of G\CTNS{} in this case is most likely an accuracy plateau, rather than a large $\cbd$.

The simplest non-Gaussian extension of the ansatz is a superposition of G\CTNS{} \cite{Karanikolaou2021GaussianContinuousTensor}.
For $\ctns=\sum\indlims\supind \supcoeff\suppsi$, the two-point function
\begin{equation}\label{eq_superposition_correlator}
    \vev{\ffop(-\kkk)\ffop(\kkk)}^{}_{\ctns}
    \propto
    \sum\nolimits_{\supind[1;2]}
    \supcoeff*[1]\supcoeff[2]\braket{\suppsi[1]}{\suppsi[2]}\brackets*{\supker*[1](\kk)+\supker[2](\kk)}\inv
\end{equation}
is still rational in $\kk^2$, hence an analogous scaling to \cref{eq_zolotarev_cbd} should hold for its degree $\order{\supnum^2\cbd}$.
For genuinely non-Gaussian \CTNS{}, on the other hand, $\vev{\ffop\ffop}^{}_{\ctns}$ is the full propagator of the local interacting Euclidean theory defined by the norm $\braket{\ctns}{\ctns}$, and need not be rational.
If that theory has a Gaussian UV fixed point, the leading UV behavior of the propagator is still governed by the quadratic part of $\ctnsfunc$, yielding the even-$z$ Lifshitz decay $\sim1/\kk^{z}$ of \cref{eq_rg_fixed_point}:
the UV mismatch persists but only model-dependent arguments or numerical studies can establish how it affects $\cbd$.
In general, interactions can also generate an anomalous UV dimension for $\ff$ --- on the QFT side, the \CTNS{} side, or both ---
and one could think of leveraging suitable non-Gaussian \CTNS{} fixed points to reproduce the correct UV of specific QFTs, although this is admittedly speculative.
\endgroup

\subsection{Entanglement entropy}\label{sec_kg_entanglement}

We showed above that G\CTNS{} approximations fail to reproduce the ultraviolet behavior of Klein-Gordon; we thus do not expect them to capture the correct scaling limit of the EE.
In light of \cref{sec_kg_nonlocal}, this mismatch can also be argued on the grounds that the G\CTNS{} represents the exact Klein-Gordon vacuum in a non-local field basis.
Since all terms in the EE \cref{eq_entropy_scaling} specific to $\spdim>1$ are regularization dependent, a first meaningful check can be carried out already in one spatial dimension.
For a 1+1$\spdim$ relativistic QFT, $g_0$ is proportional to the central charge $c$ of its UV fixed point CFT.
For a large enough bulk interval $\subsys$ \cite{Holzhey1994GeometricRenormalizedEntropy,Calabrese2004EntanglementEntropyQuantum},
\begin{equation*}
    \ent
    \sim
    \frac{c}{3} \log(\corrlen/\uv)
    \quad
    \text{when}
    \quad
    \subsys*\gg\corrlen\gg\uv
    \ ,
\end{equation*}
where $\corrlen=\mass\inv$ is the correlation length.
For a massless scalar $c=1$ and thus $g_0=1/3$.
We expect G\CTNS{} ground states to match this result up to the scale $1/\uv = \order{\pepsuv}$ at which the approximation breaks down, while
at finer spacings
the scaling is governed by the effective G\CTNS{} dispersion,
$\pepsfreq(\kk) \sim \speedoflight\kk^z$,
with some constant $\speedoflight>0$ and, for the constructions in \cref{sec_kg_dispersion}, $z=0$ or $z=2$.
The first case of an asymptotically flat dispersion implies that high-energy modes are completely localized and thus do not contribute to the entanglement: the UV fixed point is trivial.
The second case corresponds in the UV to a $z=2$ Lifshitz scalar, for which, in $1\spdim$, \cref{eq_lifshitz_g0} gives $g_0=1$.
In summary, when $\corrlen,\subsys*\gg\uv$ we anticipate:
\begin{equation}\label{eq_gctns_g0eff}
    \geff(\uv) = -\uv \odv{\ent}{\uv} \approx
    \begin{cases}
        \begin{aligned}
             & 0   &  & \text{for } \uv\ll\pepsuv^{-1} \;\;\text{and} &  & z=0              \\
             & 1/3 &  & \text{for } \pepsuv^{-1}\ll\uv\ll\corrlen     &  & \mathllap{(}z=1) \\
             & 1   &  & \text{for } \uv\ll\pepsuv^{-1} \;\;\text{and} &  & z=2
        \end{aligned}
    \end{cases}
    \!\!\!\!;
\end{equation}
and hence $g_0=\lim_{\uv\to0}\geff(\uv)\in\{0,1\}$.
In what follows we verify this claim numerically, introducing a lattice regularization.

The EE of a free lattice boson can be computed following the real-time approach of ref.~\cite{Casini2009EntanglementEntropyFree} (correlator method):
for Gaussian states,
the reduced density matrix of a subsystem $\subsys$
can be expressed solely in terms of two-point functions of the degrees of freedom situated inside $\subsys$.
For pure, time-reversal symmetric states,
its von Neumann entropy reads
\begin{equation}\label{eq_real_time_entanglement}
    \newcommand{\CC}{\mathbf{C}}
    \ent=\tr\parentheses*{
        \frac{\CC+1}{2}\log\frac{\CC+1}{2}-
        \frac{\CC-1}{2}\log \frac{\CC-1}{2}
    }
    \,, \qquad
    \CC = \sqrt{
    [\latker\inv]_{\subsys}^{}
    \latker\vphantom{]}_{\subsys}^{}
    }
    \ ,
\end{equation}
where the matrix $\latker$ is the position-space Gaussian kernel of the state,
\begin{equation}
    \frac12[\latker\inv]^{}_{\xxx[1;2]} = \vev{\ffop_{\xxx[1]}\ffop_{\xxx[2]}}
    \ , \qquad
    \frac12\latker^{}_{\xxx[1;2]} = \vev{\ppop_{\xxx[1]}\ppop_{\xxx[2]}}
    \ ,
\end{equation}
and the subscript $\subsys$ denotes restriction to $\xxx|,|[1;2]\in\subsys$.
For our G\CTNS{} approximations, $\latker_{\xxx[1;2]}$ is the discrete Fourier transform of $\kernel(\kk)$, up to lattice artifacts.
We consider a periodic chain of $\nbody$ sites, with spacing $\uv$ and length $\ir=\uv\nbody$, and set:
\begin{equation}\label{eq_lattice_kernel}
    \latker^{}_{\xx[1;2]}
    = \frac{1}{\ir}\sum\indlims\nn
    e^{i \latk (\xx|-|[1;2])}
    \kernel(\latk*)
    \ , \qquad
    \latk*
    = \frac{2}{\uv} \sin\frac{\uv\latk}{2}
    \ , \qquad
    \latk = \frac{2\pi\nn}{\ir}
    \ ,
\end{equation}
which corresponds to replacing $\nabla^2$ by the finite-difference lattice Laplacian.
The analogous substitution $\freq(\latk)\to\freq(\latk*)$ maps the Klein-Gordon Hamiltonian to the harmonic chain.
However, numerically we find that using $\latk*$ or $\latk$
does not alter $g_0$, as expected for a regularization-independent quantity.
The bulk EE \emph{scaling} is similarly insensitive to finite-size effects, justifying the introduction of a finite-volume regularization
(see also footnote~\ref{note_uv_ir_mixing}).

\begin{figure}[b]
    \includegraphics{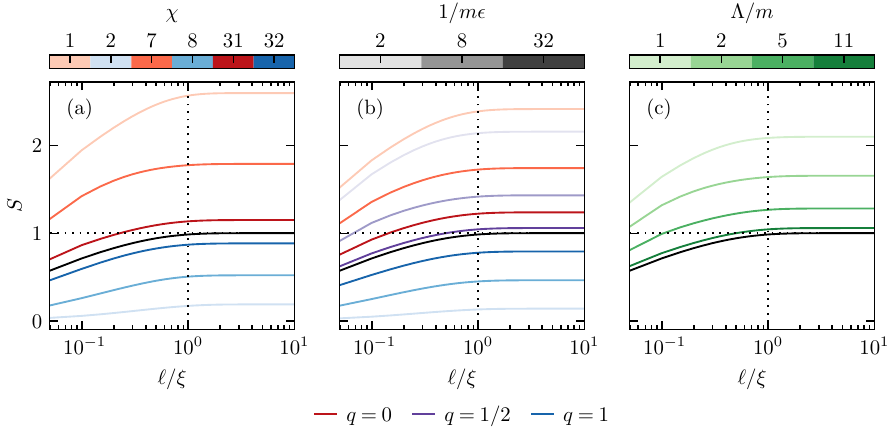}%
    \caption{%
        Lattice-regularized EE of the exact Klein-Gordon vacuum on a ring (black curve), and
        of its (a) Padé \cref{eq_pade_cf} with $\pepsuv=\order{\cbd\mass}$, (b) Trotter \cref{eq_evo_W} with $\pepsuv=\order{1/\evoeps}$, and (c) Zolotarev \cref{eq_zolo_dispersions} G\CTNS{} approximations.
        The UV scales in (a) and (b) are not directly comparable to those in (c), as in the latter the GCPEPS is uniformly accurate to $\order{10^{-5}}$ up to $\pepsuv$, whereas in (a) and (b) $\pepsuv$ is the scale at which deviations are $\order{1}$.
        In (b), $\mass\evoT=8$.
        In (c), we fix $\unirelerr=10^{-5}$ and adjust $\pepsuv$ according to $\cbd\in\{1,2,3,4\}$.
        Regulators: $\uv = e^{-3}\corrlen \approx 0.05\corrlen$, $\ir \approx 10\pi\corrlen$ (closest integer multiple of $\uv$).
        Grid lines mark $\chordlen(\corrlen) \approx \corrlen$ (vertical) and $\ent=(1/3)\log(\corrlen/\uv)=1$ (horizontal).
    }
    \subcapanchors{fig_entanglement_pade,fig_entanglement_evo,fig_entanglement_zolotarev}%
    \label{fig_entanglement}
\end{figure}
\begin{figure}
    \setlength{\belowcaptionskip}{2em}%
    \includegraphics{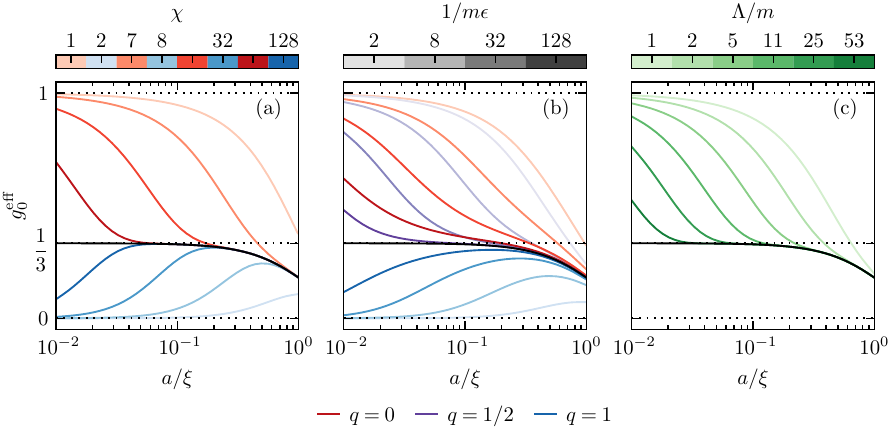}
    \caption{%
        Continuum scaling of the half-chain EE of the G\CTNS{} in \cref{fig_entanglement}, as captured by its ($\log\uv$)-derivative, matching the prediction in \cref{eq_gctns_g0eff}.
        Same parameters, extended to $\cbd\leq128$ (a), $\evonum\leq1024$ (b), and $\cbd\leq6$ (c).
        As $\uv\sim\corrlen$, discretization effects cease to be negligible \cite{Eisler2020EntanglementHamiltoniansNon}; see \cref{fig_central_charge}.
    }
    \subcapanchors{fig_entanglement_scaling_pade,fig_entanglement_scaling_evo,fig_entanglement_scaling_zolotarev}%
    \label{fig_entanglement_scaling}
\end{figure}

\begin{figure}
    \includegraphics{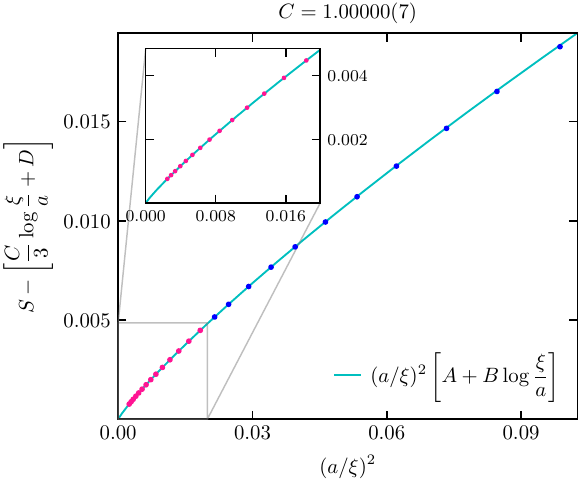}
    \caption{%
        Lattice corrections to the continuum scaling of the EE, captured by a Zolotarev G\CTNS{}
        $\zolofreq$ with $\pepsuv\approx20\pi\mass$ and $\cbd=7$, giving $\unirelerr=\order{10^{-6}}$ over the Brillouin zone of the finest lattice, $\uv=e^{-3}\corrlen$.
        The cyan curve shows
        a linear least-squares fit of $\ent(\uv)$ in \cref{eq_entropy_lattice_correction} over the $14$ smallest spacings
        (red points).
        The uncertainty on $C$ due to neglected higher-order lattice corrections is computed as the spread of the $C$ estimates obtained by successively enlarging the fitting window, until all plotted points are included.
    }
    \label{fig_central_charge}
\end{figure}

\Cref{fig_entanglement} shows the EE as given by \cref{eq_real_time_entanglement} for different bipartitions of the ring, as a function of their chord length:
\begin{equation*}
    \chordlen(\subsys*)
    = \chordlen(L-\subsys*)
    = \frac{\ir}{\pi} \sin\frac{\pi \subsys*}{\ir}
    \ .
\end{equation*}
With the regulator $\uv$ in place, G\CTNS{} representations with $\pepsfreq\sim\speedoflight\kk^z$ converge to the exact curve (black) as their UV scale $\pepsuv$ grows --- from below or above for $z=0$ and $z=2$, respectively.
On the other hand, using a symmetric bipartition as reference, \cref{fig_entanglement_scaling} clearly shows that the continuum scaling of the G\CTNS{} EE is different from the exact one and matches the prediction in \cref{eq_gctns_g0eff}.
As expected, the correct Klein-Gordon scaling still emerges in the $\pepsuv^{-1}\ll\uv\ll\corrlen$ regime, preceding the breakdown of the G\CTNS{} approximation.
Since on the lattice all momenta lie inside the Brillouin zone, $\kk\leq\pi/\uv$,
a G\CTNS{} with $\pepsuv\geq100\pi\mass$ saturates the highest lattice UV scale used in \cref{fig_entanglement_scaling}, which samples spacings $\uv\ge0.01\corrlen$.
At this $\pepsuv$, a Zolotarev G\CTNS{} with $\cbd=10$ achieves an $\order{10^{-6}}$ relative error on both the dispersion relation (for all lattice momenta, by construction)
and the half-chain EE (for all sampled $\uv$), making the latter visually indistinguishable from the exact curve in \cref{fig_entanglement_scaling}.
$\pepsuv/\mass=\order{100}$ is more than enough to recover even UV properties of the theory.
\Cref{fig_central_charge} demonstrates it with a mock extrapolation of the central charge of the conformal boson,
by fitting,
\begin{equation}\label{eq_entropy_lattice_correction}
    \ent =
    \frac{C}{3}\log\frac{\corrlen}{\uv} + D +
    (\uv/\corrlen)^2\brackets*{A+B\log\frac{\corrlen}{\uv}}
    \ ,
\end{equation}
where we include the leading quadratic lattice corrections, originating from the $\order{\uv^2\kk^4}$ term of the lattice dispersion $\freq^2((2/\uv)\sin(\uv\kk/2))$, and compatible with the exact saturated bulk EE of the harmonic chain \cite{Eisler2020EntanglementHamiltoniansNon}.
The fit is based on a Zolotarev G\CTNS{} with $\cbd=7$ and gives $C=1.00000(7)$,
confirming that, despite the UV mismatch, G\CTNS{} can accurately reproduce UV properties of Klein-Gordon theory with small bond dimension.

\section{Conclusion}\label{sec_conclusion}
\begingroup\def\CTNS{CTNS}
Gaussian continuous tensor network states (GCTNS, or GCPEPS) encode ground states of quadratic QFTs whose dispersion relations $\freq$ are rational functions of $\kk^2$, and that thus flow in the UV to free Lifshitz field theories with even dynamical exponent $z$.
We identified two canonical forms (symmetrized Jordan and tridiagonal) for G\CTNS{}.
Beyond clarifying the structure of the ansatz, canonical forms remove redundancy that would otherwise hamper variational optimization.
When the G\CTNS{} has a tridiagonal structure, we showed that $\freq$ has a natural representation as a continued fraction;
generalization to block tridiagonal matrices should be possible \cite{Huang1997AnalyticalInversionGeneral}.
In contrast with CMPS and Gaussian CMERA, which have a UV-finite entanglement entropy \cite{FrancoRubio2017EntanglementCorrelationsContinuous,Vardian2023EntanglementRenormalizationClass}, we find that the entanglement entropy is generally divergent for the (broader) G\CTNS{} class studied here, with a UV scaling matching that of Lifshitz ground states.
In one space dimension,
this manifests as a logarithm of the UV regulator multiplied by a universal coefficient, bounded by a linearly growing function of the bond dimension $\cbd$ (number of virtual fields).
We verified this numerically with some examples.
In higher dimensions, entanglement scaling in Lifshitz theories with arbitrary dynamical exponent is yet to be fully characterized, but there is ongoing progress in this direction \cite{He2017EntanglementEntropyLifshitz,MohammadiMozaffar2017EntanglementLifshitzType,Gentle2018LifshitzEntanglementEntropy,AngelRamelli2019EntanglementEntropyGeneralised,AngelRamelli2020LogarithmicNegativityQuantum,Basak2024MasslessLifshitzField,Khoshdooni2025CapacityEntanglementLifshitz}, and future studies may further elucidate the entanglement content of G\CTNS{}, at least as far as continuum scaling is concerned.

Because of their even-$z$ Lifshitz UV nature, G\CTNS{} cannot capture the short-distance properties of relativistic QFTs, such as their linear dispersion and entanglement scaling.
We studied this limitation quantitatively in the simple setting of Klein-Gordon field theory.
To this end, we first showed how to approximate free ground states by means of G\CTNS{} via rational approximants or Trotterized imaginary-time evolution.
The resulting G\CTNS{} accurately reproduce the low-energy properties of Klein-Gordon theory, while introducing a UV scale $\pepsuv$ beyond which the approximation breaks down.
This soft cutoff is raised by adding more auxiliary fields and,
using an optimal rational approximant to the relativistic dispersion
(Zolotarev's approximation of the square root),
we showed that
$\pepsuv$ grows \emph{exponentially} with $\cbd$ at fixed uniform relative accuracy across energy scales, \eg{} in correlation functions.
This makes UV extrapolations feasible at bond dimensions comparable to those used in previous non-relativistic numerical studies \cite{Karanikolaou2021GaussianContinuousTensor}, as we demonstrated by computing the central charge of the free conformal boson using as few as 7 auxiliary fields.
Beyond the analytical free-theory explorations carried out in this work, G\CTNS{} are first and foremost a variational ansatz.
After gauge fixing, a G\CTNS{} has $\sim2\cbd$ parameters and, thanks to Wick's theorem, the energy density and its full gradient can be evaluated at $\order{\cbd^3}$ linear-algebra cost \cite{Karanikolaou2021GaussianContinuousTensor}.
Combined with a small bond dimension cost of the (quadratic) UV mismatch,
this gives a positive outlook for variational G\CTNS{} optimization in interacting relativistic theories with a cutoff.
Of course, the limitations of approximating an interacting vacuum with a Gaussian ansatz persist, and are likely to be the main accuracy constraint.
In this regard, if confirmed for suitable non-Gaussian extensions of the ansatz, the small bond dimensions found here could help make their variational optimization feasible \cite{Karanikolaou2021GaussianContinuousTensor}.

Finally, at the conceptual level, the emergence of a cutoff partially defeats the purpose of introducing \emph{continuous} tensor network states in the first place \cite{Tilloy2021RelativisticContinuousMatrix}, namely offering a description of the field theory valid at \emph{all scales}.
Regrettably, G\CTNS{} approximations of relativistic quantum fields achieve the latter only for $\cbd=\infty$.
A similar problem has already been faced in one space dimension, where CMPS \cite{Verstraete2010ContinuousMatrixProduct} provide a consistent continuum TNS ansatz with finite-dimensional virtual spaces \cite{Tilloy2019ContinuousTensorNetwork}, thus allowing for variational optimization beyond Gaussianity, but are not adapted to the UV behavior of relativistic theories.
RCMPS (relativistic CMPS) \cite{Tilloy2021RelativisticContinuousMatrix} solve the problem by expanding the state in a basis of \emph{quasi-local} real-space degrees of freedom.
In this way, the correct conformal short-distance behavior is built into the ansatz.
In fact, we re-derived the RCMPS field basis from G\CTNS{} using the sole requirement of reproducing the relativistic dispersion.
While the approach can be extended to higher dimensions, crucial to the success of RCMPS is the fact that, in 1+1$\spdim$ relativistic QFTs, normal ordering is generally enough to cure all UV divergences \cite{Tilloy2021RelativisticContinuousMatrix}.
For interacting theories in higher dimensions, however, this is no longer true.
In the presence of UV divergences, variational optimization will adjust the variational parameters toward fitting shorter and shorter distances, which dominate the energy density, paradoxically degrading the accuracy at physical length scales \cite{Feynman1988DifficultiesApplyingVariational,Tilloy2021VariationalMethodRelativistic}.
Nevertheless, before RCMPS were introduced, CMPS had been successfully combined with an explicit UV cutoff to study relativistic theories, proving this is a viable approach \cite{Haegeman2010ApplyingVariationalPrinciple,Stojevic2015ConformalDataFinite}.
On the one hand, by clarifying the scales at play in the approximation, this work can therefore inform future (G)\CTNS{} investigations of interacting relativistic QFTs with a cutoff.
On the other hand, by stressing certain limitations of the ansatz, we highlight the need for new methods tailored to relativistic contexts.
\endgroup

\acknowledgments

We are grateful to Tom Shachar for his insightful contributions to the initial development of this work.
We thank Antoine Tilloy, Mari Carmen Bañuls, Erickson Tjoa, Gertian Roose and Niranjan Kumar for helpful discussions.
MR thanks the Racah Institute of Physics of the Hebrew University of Jerusalem for hospitality and partial support.
This work has received funding from the European Union under the Horizon Europe research and innovation programme through the project PASQuanS2.1 (grant agreement No. 101113690) and from the ERC project OverSign (grant agreement No. 101122583). Views and opinions expressed are however those of the author(s) only and do not necessarily reflect those of the European Union or the European Research Council. Neither the European Union nor the granting authority can be held responsible for them.
Work at MPQ is part of the Munich Quantum Valley, which is supported by the Bavarian state government with funds from the Hightech Agenda Bayern Plus.

\appendix

\section{G\CTNS{} canonical forms}\label{sec_canonical}

As anticipated in \cref{sec_ctns}, the quadratic first-order G\CTNS{} functional $\ctnsfunc$ in \cref{eq_gctns_W} is largely over-parameterized:
\begin{equation}
    \ctnsmat(\kk) = \ctnsmat{0} + \ctnsmat{1} \kk^2
    \ ,
\end{equation}
with $\ctnsmat{i}\in\mathcal{M}_{\cbd+1}(\complex)$ symmetric, contains
\begin{math}
    (\cbd+1)(\cbd+2)
\end{math}
complex parameters,
while the physical state, specified by
\begin{equation}
    \kernel(\kk) = \frac{P^{(\cbd+1)}(\kk^2)}{Q^{(\cbd)}(\kk^2)}
    \ ,
\end{equation}
has only
\begin{math}
    2\cbd+2
\end{math}.
We defined two ``canonical forms'' for G\CTNS{}: symmetrized Jordan $\vctnsmat$ and tridiagonal $\ctnsmat$,
corresponding respectively to \emph{flat} (all $\ffv$ couple directly only to $\ffv[0]$) and \emph{nested} (each $\ffv$ couples to $\ffv_{\vvind\pm1}$) physical-virtual couplings.
As we prove below, the first form can encode \emph{any} translation and rotation invariant first-order G\CTNS{} and \emph{any} valid Gaussian kernel $\kernel$ rational in $\kk^2$; the second form
has useful $\cbd\to\infty$ convergence properties and, by analytic continuation,
covers \emph{generic} G\CTNS{}, \idest{} $\ctnsmat$ or $\kernel$ picked at random.
Before showing how $\ctnsmat$ is brought into such forms,
we set the notation and recall some basic facts about Gaussian path integrals \cite{ZinnJustin2021QuantumFieldTheory} and pure Gaussian states.

For the present purposes, it is convenient to isolate the physical field $\fpp=\ffv[0]$ from the virtual multiplet
\begin{math}
    \fvv=\{\ffv\}\vvlims
\end{math}.
Accordingly, we parameterize
\begin{equation}\label{eq_gctns_blocks}
    \ctnsmat
    = \begin{bmatrix}
        \pctnsmat  & \expandafter\pvctnsmat\transp \\
        \pvctnsmat & \vctnsmat
    \end{bmatrix}
    \ ,
\end{equation}
and analogously for each $\ctnsmat{i}$, $i\in{\{0,1\}}$, with
\begin{equation}\label{eq_gctns_blocks_domains}
    \pctnsmat{i} \in \complex
    \ , \quad
    \pvctnsmat{i} \in \complex^\cbd
    \ , \quad
    \vctnsmat{i} = {\vctnsmat{i}}\transp\in\mathcal{M}_{\cbd}(\complex)
    \ .
\end{equation}

Carrying out the Gaussian integral \cref{eq_ctns}, namely
\begin{equation}\label{eq_auxiliary_integral}
    \ctns(\fpp) =
    \exp*{-\frac12\!\kint\pctnsmat(\kk)\bar\fpp\fpp}
    \int \DD\fvv \exp*{-\!\!\kint\brackets*{
            \frac12\bar\fvv\transp\vctnsmat(\kk)\fvv
            +\bar\fvv\transp\pvctnsmat(\kk)\fpp
        }
    }
    \ ,
\end{equation}
one readily obtains \cref{eq_gctns_kernel}, namely
\begin{equation}\label{eq_gctns_kernel_blocks}
    \kernel = \pctnsmat - \expandafter\pvctnsmat\transp\expandafter\vctnsmat\inv\pvctnsmat
    \ .
\end{equation}
The path integral formally (\idest{}, mode-by-mode)
converges iff the Hermitian part $\re\vctnsmat(\kk)$ of $\vctnsmat(\kk)$ is positive definite (as a real matrix) $\forall\kk$, \idest{} iff
\begin{equation}\label{eq_gctns_convergence}
    \re\vctnsmat{0}\succ0
    \;\;\text{and}\;\;
    \re\vctnsmat{1}\succeq0
    \ ,
\end{equation}
implying that, as a function on the complex $\kk^2$ plane, $\kernel$ has no poles for (physical) $\kk^2\geq0$.
\begin{proof}
    Poles of $\kernel$ are contained in the zeros of $\det\vctnsmat$
    and, for $\vctnsmat$ complex symmetric,
    $\re\vctnsmat\succ0 \implies\vctnsmat$ is non-singular.
\end{proof}

Since
\begin{equation}
    \norm{\ctns}^2
    = \int \DD\fpp\: \abs{\ctns(\fpp)}^2
    = \int \DD\fpp\: \exp*{-\kint\fpp(-\kkk)\re\kernel(\kk)\fpp(\kkk)}
    \ ,
\end{equation}
$\ctns$ is formally normalizable
iff, moreover, $\re\kernel(\kk)>0\;\forall\kk$ --- often relaxed by allowing a zero at $\kk=0$, with an implicit regulator, to accommodate massless fields.
The same requirement in the conjugate momentum representation, $\ctns(\pp)$, where $\kernel\to\kernel\inv$, reads $\re\kernel\inv=\re\kernel/\abs{\kernel}^2>0$,
recovering the no-physical-pole condition implied by \cref{eq_gctns_convergence}, which is thus intrinsic to the state, irrespective of its G\CTNS{} realization.
The normalizability condition,
\begin{equation}\label{eq_normalizable}
    \re\kernel(\kk) > 0
    \quad\text{and}\quad
    \kernel(\kk) < \infty
    \quad
    \forall\kk^2\geq0
    \ ,
\end{equation}
can be formulated in a representation-independent way by demanding a locally finite (in~$\kk$) excitation density over any regular Fock vacuum (a necessary condition for Shale's criterion).
Equivalently, recalling \cref{eq_gctns_correlators}, it enforces finite field and conjugate momentum fluctuations for every mode (no infinite squeezing).

As a corollary of the no-pole condition, the parent Hamiltonian \cref{eq_parent} with $\freq=\re\kernel$ is quasi-local: real-space couplings in $\pham$ decay exponentially with distance, with a decay length set by the inverse distance of the closest complex pole to the real axis in the $\kk$ plane.
Finally, by a property of the Schur complement \cite[Proposition 16.1]{Gallier2011SchurComplementsApplications},
\begin{equation}\label{eq_normalizable_state}
    \re\vctnsmat\succ0
    \;\;\text{and}\;\;
    \re\kernel>0
    \;\iff\;
    \re\ctnsmat\succ0
    \ .
\end{equation}
In the following we explicitly enforce \cref{eq_gctns_convergence} but not \cref{eq_normalizable_state}.

The G\CTNS{} gauge freedom, by which we aim to put $\ctnsmat$ in canonical form, consists of transformations that leave $\fpp$ and $\kernel$ unchanged and respect the G\CTNS{} structure.
These correspond to local linear auxiliary-field redefinitions,
\begin{equation}\label{eq_gctns_transf}
    \fvv \to \transmat\fvv + \transshift\fpp
    \ ,
\end{equation}
with $\kkk$-independent
\begin{math}
    \transmat\in\GLN(\cbd,\complex)
\end{math},
\begin{math}
    \transshift
    \in\complex^\cbd
\end{math}.
The transformation \cref{eq_gctns_transf} clearly preserves
the path-integral measure $\DD\fvv(\kkk)$ (up to a constant) and Gaussianity of \cref{eq_auxiliary_integral},
amounting to
\begin{equation}\label{eq_gctns_transf_W}
    \vctnsmat\to\transmat\transp\vctnsmat\transmat
    \ , \quad
    \pvctnsmat\to\transmat\transp(\pvctnsmat+\vctnsmat\transshift)
    \ , \quad
    \pctnsmat\to\pctnsmat+\transshift\transp\vctnsmat\transshift+2\transshift\transp\pvctnsmat
    \ .
\end{equation}
However, when $\transmat$ or $\transshift$
have complex entries, one should in principle regard each $\fvv(\kkk)$ integral after the transformation as a contour integral over a tilted and shifted real hyperplane in $\complex^{\cbd}$.
Nevertheless, provided the transformed virtual block
still satisfies $\re\transmat\transp\vctnsmat\transmat\succ0$, the contour can be deformed back to real $\fvv$.

\subsection{Symmetrized Jordan normal form}\label{sec_symm_jordan}
\begingroup
\def\ee{\VField{e}}
\def\ss{\VField{\tilde{v}}}

Any G\CTNS{} can be written using a symmetrized Jordan \cite{Dokovic2004TridiagonalNormalForms} block diagonal $\vctnsmat$, as follows:
\begin{equation}\label{eq_canonical_jordan}
    \def\Uimproper{\idmat_{\polecbd_0\!} {+} \kk^2\nilmat_{\polecbd_0\!}}
    \def\Uproper#1{
        \sjndia_{#1}(\kk) \idmat_{\polecbd_{#1}\!} {+} \nilmat_{\polecbd_{#1}}
    }
    \def\arraystretch{1.3}
    \ctnsmat(\kk) = \begin{bmatrix}
        a{+}b\kk^2      & \kk^2\vcpl\transp_0 & \vcpl\transp_1 & \vcpl\transp_2 & \cdots \\
        \kk^2\vcpl^{}_0 & \Uimproper          &                &                &        \\
        \vcpl^{}_1      &                     & \Uproper{1}    &                &        \\
        \vcpl^{}_2      &                     &                & \Uproper{2}    &        \\
        \vdots          &                     &                &                & \ddots
    \end{bmatrix}
    \,,
\end{equation}
where
$\nilmat_{\poleord[-1]}$ is the \emph{nilpotent symmetrized Jordan block} of size $\poleord[-1]$ defined in ref.~\cite{Dokovic2004TridiagonalNormalForms},
and
\begin{equation}\label{eq_ugly_positivity}
    \sjndia(\kk) = \sjnmass + \frac{\abs{\sjnmass}}{\sjnmass}\kk^2
    \ , \qquad
    \re \sjnmass > 0
    \ .
\end{equation}
\Cref{eq_gctns_kernel_blocks} then gives:
\begin{equation}\label{eq_gctns_kernel_general}
    \kernel(\kk) =
    a + b\kk^2
    + {\def\polenum{\Ind{\Primed}{0}{}{}}\sum\indlims{\poleord} \poleres \kk^{2+2\poleord}}
    + \sum\indlims{\polenum}\sum\indlims{\poleord}\frac{\poleres}{(\polemass+\kk^2)_{\mathstrut}^{\poleord}}
    \ ,
\end{equation}
with
\begin{equation}\label{eq_gctns_kernel_coeffs}
    \polemass = \frac{\sjnmass^2}{\abs{\sjnmass}}
    \ , \qquad
    \poleres = -\frac{\sjnmass}{\abs{\sjnmass}} \polecpl\transp\nilmat_{\polecbd}^{\poleord-1}\polecpl^{\vphup}
    \ ,\qquad
    \sjnmass_0=1
    \ .
\end{equation}
Given $\kernel$ above, minimal $\cbd$ further implies
\begin{equation}\label{eq_minimal_chi}
    \polemass \neq \polemass[2]\;\;\forall\:\polenum|\neq|[1;2]
    \ , \qquad
    c_{\polenum,\poleord[-1]} \neq 0
    \ .
\end{equation}
With these conditions, the parameterization \cref{eq_canonical_jordan} is \emph{unique} up to the ordering of the diagonal blocks \cite{Dokovic2004TridiagonalNormalForms} and a sign choice for each $\polecpl$;
similarly, \cref{eq_gctns_kernel_general} is a unique parameterization of a general rational function of $\kk^2$ with
numerator of degree $p\leq\cbd+1$ and denominator of degree $q=\cbd-\polecbd_0\leq\cbd$ \cite{EMS2025RationalFunction}.
Within the minimal-$\cbd$ assumption, at least one of $p$ and $q$ must be maximal.
Relaxing that assumption, one concludes that first-derivative G\CTNS{} with $\cbd$ auxiliary fields {correspond} to the subclass of Gaussian states whose kernel is a rational function of $\kk^2$ of (inclusive) degree $\kerdeg$.

A few comments are due.
First, as expected, the number of parameters in $\ctnsmat$ and $\kernel$ matches exactly.
Second, the parameterization \cref{eq_ugly_positivity} is to ensure $\re\vctnsmat\succ0$, so that the Gaussian integral formally converges; however, analytically continuing \cref{eq_gctns_kernel_blocks} to arbitrary nonsingular $\vctnsmat$ we can obtain the same $\kernel$ as above simply setting
\begin{equation}
    \sjndia(\kk) = \polemass + \kk^2
    \ , \qquad
    \polemass \not\in (-\infty,0]
    \ ,
\end{equation}
with the understanding that the path-integral contour is rotated accordingly.
Finally, we comment on \emph{typicality}:
generic rational functions $\kernel$ have maximal degree and only simple poles;
symmetrically, generic complex symmetric matrices $\vctnsmat{0}, \vctnsmat{1}$ are diagonalizable with distinct (\idest{}, simple), nonzero eigenvalues.
Both conditions correspond to $\polecbd_0=0$ and $\polecbd=1$ for all $\polenum\geq1$, namely a diagonal virtual block
\begin{equation}
    \vctnsmat = \diag\parentheses*{
        \polemass_1 + \kk^2,\:
        \dots,\:
        \polemass_\cbd + \kk^2
    }
    \ ,
\end{equation}
recovering \cref{eq_rational_simple_poles} of the main text, \idest{} the ansatz of ref.~\cite{Karanikolaou2021GaussianContinuousTensor}.

\begin{proof}
    That $\ctnsmat$ in \cref{eq_canonical_jordan} yields $\kernel$ in \cref{eq_gctns_kernel_general} is seen immediately by inverting $\vctnsmat$ blockwise, using the identity:
    \begin{equation}
        \def\II{\idmat_{\blockdim}}
        \def\NN{\nilmat_{\blockdim}}
        \brackets[\big]{\alpha\II^{}+\beta\NN^{}}\inv
        = \sum\indlims{\poleord}^\blockdim\alpha^{-\poleord}\beta^{\poleord-1}\NN^{\poleord-1}
        \ , \quad
        \NN^0 \defeq \II^{}
        \ .
    \end{equation}
    We already know that G\CTNS{} correspond to rational kernels of degree $\kerdeg$.
    Since \cref{eq_gctns_kernel_general} parametrizes the most general such function, it is enough to show that \cref{eq_gctns_kernel_coeffs} can be solved for arbitrary $\{\polemass\}$ and $\{\poleres\}$.
    It then follows that $\ctnsmat$ can be written as above and that \cref{eq_minimal_chi} is not restrictive.

    The equation for $\polemass$ is trivial, and
    the constraint
    \begin{math}
        \re\sjnmass>0 \implies \arg(\polemass)\neq\pi
    \end{math}
    only excludes poles in the $\kk^2\geq0$ axis, as per \cref{eq_normalizable}.
    The equations for $\poleres$ in different blocks $\polenum$ are independent; thus it suffices to consider a single block of arbitrary size $\blockdim$.
    For $\blockdim=1$, obviously there are two solutions for all $c\neq0$.
    For larger $\blockdim$,
    recalling $\nilmat=\nilmat_{\blockdim}$ is symmetric and cyclic \cite{Dokovic2004TridiagonalNormalForms}, by the Jacobian criterion \cite{Beecken2013AlgebraicIndependenceBlackbox} one concludes that the equations
    \begin{equation}\label{eq_gctns_kernel_coeffs_block}
        \csol[1]=\vcpl\transp\nilmat^{\yind-1}\vcpl
    \end{equation}
    are algebraically independent.
    Then, invoking some notions of algebraic geometry \cite{Lowther2011AnswerWhenAre,Milne2024AlgebraicGeometryV610}, it can be argued that \cref{eq_gctns_kernel_coeffs_block} can be solved for $\vcpl$ at least for \emph{generic} $\csol\in\complex^\blockdim$.
    Below we give a more direct proof, which shows that the measure-zero exceptional set where \cref{eq_gctns_kernel_coeffs_block} does not have a solution lies inside the $\csol[-1]=0$ subspace, where $\cbd$ is not minimal.

    $\nilmat$ is similar to the usual Jordan nilpotent \cite{Dokovic2004TridiagonalNormalForms}:
    \begin{equation}
        \Psim\inv\nilmat\Psim=\Jnil
        \ , \quad
        \Jnil_{\yind[2;3]}=\delta_{\yind[2]+1\INDSEP\yind[3]}
        \ .
    \end{equation}
    Given a cyclic vector $\vcplcyc$ for $\nilmat$, the similarity matrix $\Psim$ can be constructed using as columns
    the Jordan chain
    \begin{equation}
        \ee[-1]=\vcplcyc
        \ , \quad
        \ee_{\yind-1}=\nilmat\ee[1]
        \ , \quad
        \yind=\yind|,\dots,|[0;-1]-1
        \ .
    \end{equation}
    In the $\ss=\Psim\inv\vcpl$ basis, \cref{eq_gctns_kernel_coeffs_block} reads
    \begin{equation}\label{eq_gctns_kernel_coeffs_gram}
        \csol[1]=\ss*\gram\Jnil^{\yind-1}\ss
        \ .
    \end{equation}
    The Gram matrix $\gram$
    is lower anti-triangular (by nilpotency of $\nilmat$),
    and $\Jnil$ acts on it by shifting its columns to the right.
    Therefore, \cref{eq_gctns_kernel_coeffs_gram} defines a triangular system of quadratic equations, $\csol[1]=f_{\yind}^{}(\ss[1],\dots,\ss[-1])$, which admits a solution if all pivots are nonzero.
    As we show below, $\gram$ can be taken to be
    the exchange matrix,
    \begin{math}
        (\gram)_{\yind[2;3]}=\delta_{\yind|+|[2;3]\INDSEP\blockdim+1}
    \end{math}.
    Then \cref{eq_gctns_kernel_coeffs_gram} takes the simple form,
    \begin{equation}
        \csol[1]=\;\sum_{\mathclap{\yind|+|[2;3]=\yind|+|[-1;1]}}\;\ss[2]\ss[3]
        \ ,
    \end{equation}
    and we can iteratively solve for $\ss$ (and thus $\vcpl$):
    \begin{equation}\label{eq_jordan_residues_sol}
        \csol[-1] = 2\ss[-1]\ss[-1] \neq 0
        \,, \quad
        \csol_{\blockdim-1} = 2\ss_{\blockdim-1}\ss[-1]
        \,, \quad
        \dots
        \,, \quad
        \csol[1]
        = 2\ss[1]\ss[-1]
        + \sum_{\mathclap{\yind|<|[1;4;-1]-\yind}} \ss[4]\ss_{\yind|-|[-1;4]}
        \,, \quad
        \dots
    \end{equation}
    The equation for $\ss[-1]$ is quadratic while all the others are linear; thus, overall, there are exactly two solutions: $\pm\ss_\star$.

    All that is left to show is that we can choose $\vcplcyc$ cyclic such that $\gram$ is the exchange matrix.
    In general,
    \begin{equation}
        (\gram)^{}_{\yind[2;3]}
        = \ee*[2]\ee[3]
        = \vcplcyc\transp\nilmat^{2\yind|-|[-1;2;3]}\vcplcyc
        \eqdef \hank_{\yind|+|[2;3]-\blockdim}
        \ ,
    \end{equation}
    is Hankel and lower anti-triangular ($\hank=0$ for $\yind<1$).
    If $\vcplcyc$ is cyclic, so is $Q(\nilmat)\vcplcyc$ with
    \begin{equation}
        Q(z)=\qpoly_0+\qpoly_1z+\dots+\qpoly_{\blockdim-1}z^{\blockdim-1}
        \ , \quad
        \qpoly_0\neq0
        \ .
    \end{equation}
    The transformation acts transitively on the moments $\{\hank\}\indlims\yind$, and
    \begin{equation}\label{eq_hakel_to_exchange}
        Q\cdot \hank
        = \vcplcyc\transp\nilmat^{\yind|-|[-1;1]}Q^2(\nilmat)\vcplcyc
        = \;\sum_{\mathclap{\yind|+|[2;3]\leq\yind-1}}\;
        \qpoly[2]\qpoly[3] \hank_{\yind|-|[1;2;3]}
        \overset!= \delta_{\yind|,|[1;0]}
    \end{equation}
    defines an upper triangular system which can be solved iteratively, starting from $\yind=0$,
    provided the pivots are nonzero, namely
    $\hank[0]\neq0$.
    This is always guaranteed because $\nilmat^{\blockdim-1}$ is rank-1 symmetric, hence
    \begin{math}
        \nilmat^{\blockdim-1} = \nilvec\nilvec\transp
    \end{math}
    for some $\nilvec\neq0$, thus
    \begin{equation}
        \hank[0]
        = (\nilvec\transp\vcplcyc)^2 = 0
        \;\implies\;
        \nilmat^{\blockdim-1}\vcplcyc = 0
        \ ,
    \end{equation}
    which is in contradiction with $\vcplcyc$ being cyclic.
\end{proof}
Notice that the existence of a solution to \cref{eq_hakel_to_exchange} is equivalent (lifting the action of $Q$ from $\vcpl$-space to $\ss$-space) to the statement that a non-degenerate anti-triangular Hankel matrix is congruent to the exchange matrix through a triangular Toeplitz matrix $Q(\Jnil)$.

Given a rational function in the form \cref{eq_gctns_kernel_general}, the associated G\CTNS{} matrix is thus obtained solving \cref{eq_gctns_kernel_coeffs} via \cref{eq_jordan_residues_sol}
(given the closed-form expression for $\nilmat$ in ref.~\cite{Dokovic2004TridiagonalNormalForms}, it should even be possible to compute $\Psim$ explicitly).
Conversely, a given $\ctnsmat$ can be put in the form of \cref{eq_canonical_jordan} via the transformation \cref{eq_gctns_transf} as follows:
\begin{enumerate}
    \item\label{jordan_steps_diagonal}
          Diagonalize $\vctnsmat{0}\to\idmat_{\cbd}$ by congruence with an invertible $\transmat$, then
          eliminate $\pvctnsmat{0}\to0$ by a shift with $\transshift=-\transmat\transp\pvctnsmat{0}$;
          both operations are possible because $\vctnsmat{0}$ is non-singular.
    \item\label{jordan_steps_jordan}
          Put $\vctnsmat{1}$ in symmetrized Jordan normal form \cite{Dokovic2004TridiagonalNormalForms}
          by a complex orthogonal rotation $\transmat$
          (residual gauge freedom);
          equivalently, compute the usual Jordan decomposition of $\vctnsmat{1}$ and replace each Jordan block by a symmetrized one with the same eigenvalue, as given in closed form in ref.~\cite{Dokovic2004TridiagonalNormalForms}.
\end{enumerate}
The virtual block then takes a block diagonal form:
\begin{equation}\label{eq_canonical_jordan_1}
    \vctnsmat = \idmat_\cbd + \bigoplus\nolimits_{\polenum} \brackets[\big]{
        \lambda_{\polenum}\idmat_{\poleord[-1]} + \nilmat_{\poleord[-1]}
    }\kk^2
    \ .
\end{equation}
Permute the blocks so that any null eigenvalues of $\vctnsmat{1}$ appear first; then repeat steps~\labelcref{jordan_steps_diagonal,jordan_steps_jordan} above on the remaining $\polenum\geq\ell$ blocks, where $\vctnsmat{1}$ is non-singular --- this time with the roles of $\vctnsmat{0}\!, \pvctnsmat{0}$ and $\vctnsmat{1}\!, \pvctnsmat{1}$ exchanged.
Then,
\begin{equation}
    \def\Uimproper#1{
        \brackets[\big]{\idmat_{\polecbd_{#1}} + \kk^2\nilmat_{\polecbd_{#1}}}
    }
    \def\Uproper#1{
        \brackets[\big]{(\polemass_{#1}+\kk^2) \idmat_{\polecbd_{#1}\!} + \nilmat_{\polecbd_{#1}}}
    }
    \vctnsmat =
    \Uimproper{0} \oplus
    \cdots
    \oplus \Uproper{\ell} \oplus
    \cdots
    \ ;
\end{equation}
\idest{} the same structure as \cref{eq_canonical_jordan}
except for, possibly,
\begin{enumerate*}
    \item degenerate blocks ($\ell>1$, geometric multiplicity $\operatorname{gm}(\polemass)>1$), and
    \item $\re\polemass\not>0$
\end{enumerate*}.
Provided $\polemass$ is not real negative, positive definiteness can be enforced by rotating the sub-multiplet $\fvv_{\polenum}$ corresponding to the $\polenum$th block by a global phase
\begin{equation}
    \fvv_{\polenum} \to
    \sqrt{\frac
        {\abs{\polemass}}
        {\,\polemass}
    }
    \fvv_{\polenum}
    \qquad\text{so that}\qquad
    \polemass + \kk^2 \to \sjndia(\kk)
    \ ,
\end{equation}
with $\sjndia$ as defined in \cref{eq_ugly_positivity,eq_gctns_kernel_coeffs}.
Finally, blocks $\polenum|\neq|[1;2]$ with degenerate eigenvalues $\polemass=\polemass[2]$ and $\polecbd\geq\polecbd[2]$ are merged by computing the associated residues $\poleres,\poleres[2]$ via \cref{eq_gctns_kernel_coeffs}, then solving \cref{eq_gctns_kernel_coeffs} with $\poleres\to \poleres+\poleres_{\polenum[2]\INDSEP\poleord}$ (padding the latter with zeros if ${\polecbd}>\polecbd[2]$) to obtain a new
$\polecpl$, and finally dropping the $\polenum[2]$ block.
\endgroup

\subsection{Tridiagonal form \& continued fractions}\label{sec_tridiagonal}
\emph{Generic} G\CTNS{} matrices can, by means of \cref{eq_gctns_transf}, be brought to tridiagonal form with constant off-diagonals:
\begin{equation}\label{eq_canonical_tridia}
    \ctnsmat =
    \begin{bmatrix}
        \tridia_0 & i         &         &                \\
        i         & \tridia_1 & \sddots &                \\
                  & \sddots   & \sddots & i              \\
                  &           & i       & \tridia_{\cbd} \\[3pt]
    \end{bmatrix}
    ,
\end{equation}
which depends on $2\cbd+2$ complex parameters $\tridiac,\tridiak\in\complex$ via
\begin{equation}
    \tridia(\kk) = \tridiak\kk^2 + \tridiac
    \ , \qquad
    \tridiak\neq0\;\;\text{for}\;\;\vvind|=|[1;0],\dots,\vvind[-1]
    \ .
\end{equation}
\Cref{eq_gctns_kernel_blocks} then gives
\begin{equation}\label{eq_canonical_tridiagonal_cf}
    \kernel
    = \tridia[0] + \Uinv
    = \tridia[0] + \cfrac{1}{
        \tridia_1 + \cfrac{1}{
            \cfracdots + \cfrac{1}{
                \tridia_{\cbd}}}}
    \ ,
\end{equation}
which \emph{uniquely} encodes a \emph{generic} rational function of degree $\kerdeg$.
Its proper part, \idest{} the nontrivial GCPEPS contribution $\Uinv$, is a (truncated) Jacobi-type continued fraction, or \emph{J-fraction} for short \cite{Wall1948AnalyticTheoryContinued}.
\begin{proof}
    That $\ctnsmat$ in \cref{eq_canonical_tridia} gives rise to $\kernel$ in \cref{eq_canonical_tridiagonal_cf} was already shown in \cref{sec_ctns}, where we also argued that the latter encodes generic rational functions \cite[Problem~3.11]{Sauer2021RationalFunctionsAs} --- more precisely, those in \cite[Theorem~41.1]{Wall1948AnalyticTheoryContinued}.
    Uniqueness of the J-fraction follows from that of the quotient and remainder of polynomial division at each step of the Euclidean algorithm, and from $\tridiak[-1]\neq0$, which removes ``tail ambiguities''; an explicit proof is given in \cite[pp.~161--162]{Wall1948AnalyticTheoryContinued}.

    \newcommand{\oldmat}{\mathbf{X}}
    \newcommand{\newmat}{\mathbf{Y}}
    \newcommand{\OO}{\mathbf{O}}
    \newcommand{\oldvec}{\mathbf{a}}
    \def\tridia{b}
    \def\trioff{a}

    To put a generic $\ctnsmat$ in the form of \cref{eq_canonical_tridia}, we need to
    \begin{enumerate*}
        \item diagonalize $\ctnsmat{1}$ and
        \item bring $\ctnsmat{0}$ to tridiagonal form.
    \end{enumerate*}
    We may then assume all off-diagonal elements are nonzero and can thus be rescaled to $i$ by renormalizing $\ffv$;
    otherwise, the continued fraction terminates early and some virtual fields decouple, implying that $\cbd$ is not minimal.
    Diagonalization of $\ctnsmat{1}$ works as in step~\labelcref{jordan_steps_diagonal} of the prescription on \cpageref{jordan_steps_diagonal}, here applied to $\pvctnsmat{1}$ and $\vctnsmat{1}$ (generically non-singular);
    in particular, $\vctnsmat{1}\to\idmat$.
    To tridiagonalize $\ctnsmat{0}$ by a complex orthogonal rotation of the virtual fields (residual gauge),
    observe that, under congruence by $\idmat_1\oplus\OO$ orthogonal, a symmetric matrix $\oldmat$ transforms as
    \begin{equation}
        \oldmat = \begin{bmatrix}
            \tridia & \oldvec\transp \\
            \oldvec & \newmat
        \end{bmatrix}
        \;\longmapsto\;
        \oldmat' = \begin{bmatrix}
            b                 & \oldvec\transp\OO \\
            \OO\transp\oldvec & \newmat'
        \end{bmatrix}
        \ ,\qquad
        \newmat' = \OO\transp\newmat\OO
        \ .
    \end{equation}
    Generically $\oldvec\transp\oldvec\neq0$, in which case $\OO$ can be chosen such that $\oldvec\transp\OO=(\trioff,0,\dots,0)$,
    with $\trioff=\pm\sqrt{\oldvec\transp\oldvec}$
    by orthogonality.
    The procedure is then repeated for $\oldmat\leftarrow\newmat'$, iterating until $\newmat'$ is one-dimensional.
\end{proof}

Although here we focused on the generic case,
the above prescription can be adapted to singular $\vctnsmat{1}$ --- \idest{}, kernels with sub-maximal denominator degree --- by swapping the roles of $\ctnsmat{0}$ and $\ctnsmat{1}$.
The off-diagonals of $\ctnsmat$ in \cref{eq_canonical_tridia} and the numerators of the continued fraction \cref{eq_canonical_tridiagonal_cf} then acquire a $\kk^2$ and $\kk^4$ factor, respectively.

Unlike in \cref{eq_canonical_jordan}, $\re\vctnsmat$ is \emph{not necessarily positive definite} for $\ctnsmat$ in \cref{eq_canonical_tridia}.
Unless
\begin{equation}\label{eq_tridia_positivity_naive}
    \re\tridiac,\,\re\tridiak\relax>0
    \,,\quad
    \vvind=\vvind|,\dots,|[0;-1]
    \ ,
\end{equation}
the auxiliary-field path integral \cref{eq_auxiliary_integral} converges only if regarded as an appropriately tilted contour integral.\footnote{We stress that the continued fraction expansion of $\Uinv$ in \cref{eq_canonical_tridiagonal_cf} holds regardless \cite{Usmani1994InversionJacobisTridiagonal}.}
This issue is partially cured by introducing a slightly less rigid ansatz:
relaxing $\ctnsmat_{\vvind,\vvind+1}=i$ gives the additional gauge freedom,
\begin{equation}\label{eq_tridia_phase}
    \ffv \to \sqrt{\trioff}\ffv
    \ , \qquad
    \trioff[0]= 1
    \ ,
\end{equation}
which corresponds to an equivalence transformation of continued fractions:
\begin{equation}
    \tridia_0 + \cfrac{1}{
        \tridia_1 + \cfrac{1}{
            \cfracdots + \cfrac{1}{
                \tridia_{\cbd}}}}
    \:=\:
    \tridia_0 + \cfrac{\trioff_1}{
        \trioff_1\tridia_1 + \cfrac{\trioff_1\trioff_2}{
            \cfracdots + \cfrac{\trioff_{\cbd-1}\trioff_{\cbd}}{
                \trioff_{\cbd}\tridia_{\cbd}}}}
    \;\;.
\end{equation}
With it, some cases where \cref{eq_tridia_positivity_naive} does not hold can be brought into a form where $\re\vctnsmat\succ0$, and thus $\fvv$ is real,
at the expense of a redundant parameterization.
However, this does not apply to all J-fractions, not even generically.
\begin{proof}
    Since $\trioff\relax>0$ cannot change the signature of $\re\vctnsmat$ (Sylvester's law of inertia), it is enough to consider phase rotations,
    $\trioff=e^{i\triarg}$.
    By these,
    whenever
    $\tridiac\neq0$ and $\tridiak,\tridiac$ are not in counter-phase
    (\idest{} generically),
    diagonal entries $\re\trioff\tridia$
    can be made positive,
    ensuring $\re\vctnsmat{1}\succ0$ but not necessarily
    $\re\vctnsmat{0}\succ0$ (because of the off-diagonals);
    and one can construct examples, neither generic nor exceptional, where the latter can be enforced and vice versa.
    Let $\triint$ be the interval of angles $\triarg$ compatible with positive diagonal entries:
    \begin{equation}
        \triint = \{
        \triarg \in [0,2\pi)
        :
        \re\trioff\tridiak,\re\trioff\tridiac\relax>0 \}
        \ ,
    \end{equation}
    and let $\triint[1;2]=\triint+\triint[2] \pmod{2\pi}$;
    notice that $\triint$ shrinks as $\tridiak,\tridiac$ approach counter-phase.
    If $0\in \triint_{\vvind,\vvind+1}\;\forall\vvind$,
    off-diagonal $\vctnsmat{0}$ entries can be kept purely imaginary, ensuring $\re\vctnsmat{0}\succ0$ regardless of \cref{eq_tridia_positivity_naive}.
    Conversely,
    if $\triint_{\vvind,\vvind+1}\subset(\pi-\epsilon,\pi+\epsilon)$ for some $\vvind$, with
    \begin{equation}
        0<\epsilon<\pi
        \ , \qquad
        \abs{\tridiac}\abs{\tridiac>}<\cos^2(\epsilon/2)
        \ ,
    \end{equation}
    then the $(\vvind,\vvind{+}1)$ principal minor of $\re\vctnsmat{0}$ is negative:
    \begin{equation}
        \re(\trioff\tridiac)\re(\trioff>\tridiac>) -
        \brackets*{\re(i\sqrt{\trioff\trioff>})}^2
        < 0
        \ ,
    \end{equation}
    and thus $\re\vctnsmat{0}$ cannot be made positive-definite.
\end{proof}

Choosing $\trioff=\tridiak^{-1}$ in \cref{eq_tridia_phase}
gives $\vctnsmat{1}\to\idmat$,
corresponding to monic denominators in
\begin{equation}
    \Uinv(\kk) =
    \cfrac{1}{
        \kk^2+\tridiac_1/\tridiak_1-\cfrac{-1/\tridiak_1\tridiak_2}{
            \cfracdots-\cfrac{-1/\tridiak_{\cbd-1}\tridiak_{\cbd}}{
                \kk^2 + \tridiac_{\cbd}/\tridiak_{\cbd}}}}
    \;\;.
\end{equation}
Within this ansatz,
J-fractions realizable with $\re\vctnsmat\succ0$ correspond to generic (truncated) \emph{positive-definite J-fractions} \cite{Wall1948AnalyticTheoryContinued}.
This class
encompasses many useful continued fraction expansions with nice and extensively studied convergence properties \cite{Wall1948AnalyticTheoryContinued}.
It partially overlaps with \cref{eq_tridia_positivity_naive}, but neither is a subset of the other.

\begin{proof}
    \let\fvind\triind
    \def\tridia{\vField{b}}
    \def\trioff{\vField{a}}
    We first recall the definition of a positive-definite continued fraction, in the notation of ref.~\cite{Wall1948AnalyticTheoryContinued}, then show how it (generically) relates to $\re\vctnsmat\succ0$ by a $\pi/2$ rotation of the complex plane.
    Consider the J-fraction
    \begin{equation}\label{eq_jfrac_monic}
        \Jinv(z) = \cfrac{1}{
            z+\tridia_1-\cfrac{\trioff_2^2}{
                z+\tridia_2-\cfrac{
                    \;\trioff_3^2\;}{\ddots}}
        }
        \;\;,
    \end{equation}
    formally originating from the expansion of the leading coefficient of the reciprocal of the infinite J-matrix \cite{Wall1948AnalyticTheoryContinued}
    \begin{equation}\label{eq_jmat_monic}
        \Jmat(z) = z\idmat + \Jmat^{(0)}
        \ ,\qquad
        \Jmat^{(0)} =
        \begin{bmatrix}
            \tridia_1 & \trioff_2 &           &         \\
            \trioff_2 & \tridia_2 & \trioff_3 &         \\
                      & \trioff_3 & \tridia_3 & \sddots \\
                      &           & \sddots   & \sddots \\
        \end{bmatrix}
        \ .
    \end{equation}
    Following Wall \cite[Definition~16.1]{Wall1948AnalyticTheoryContinued}, $\Jinv$ is positive-definite when the real quadratic form $\im\Jmat^{(0)}$ is non-negative on eventually zero sequences:
    \begin{equation}\label{eq_positive_cf}
        \sum\indlims{\triind}\im(\tridia)\trixi^2 +
        2\sum\indlims{\triind}_2\im(\trioff)\trixi<\trixi \geq 0
        \qquad\forall\;\xi\in\reals^{\triord},\;\triord=1,2,\dots
        \ ;
    \end{equation}
    equivalently, when $\im\Jmat_N^{(0)}\succeq0$ for all leading principal submatrices $\Jmat_N^{(0)}$ of $\Jmat^{(0)}$.
    Among other things, this implies that if $\Jinv$ converges for one point with $\im z>0$, then it converges to an analytic function on the whole upper half plane \cite[p.~110]{Wall1948AnalyticTheoryContinued}, and all its approximants (truncations) $(\Jmat_N\inv)^{}_{11}$ have no poles for $\im z>0$ \cite[Theorem~16.1]{Wall1948AnalyticTheoryContinued}.

    Suppose $\Jmat$ defines a positive-definite J-fraction.
    Making the replacement $\Jmat\to i\Jmat$ or, equivalently,
    \begin{equation}
        z \to iz
        \ , \quad
        \trioff \to i\trioff
        \ , \quad
        \tridia \to i\tridia
        \ ,
    \end{equation}
    \cref{eq_positive_cf} and the convergence results that follow hold with $\im\to\re$.
    Therefore, if $i\vctnsmat(\kk)=\Jmat_{\cbd}(i\kk^2)$, then $\re\vctnsmat\succeq0$, and thus generically $\re\vctnsmat\succ0$, whenever $\re\kk^2\geq0$ and in particular for $\kk^2\geq0$.
\end{proof}

\section{Higher-order derivative \CTNS{}}\label{sec_higher_order}
We briefly show how, by introducing additional auxiliary fields, a \CTNS{} with $\dvord$-order derivatives can be recast as a first-order one.
\Cref{sec_symm_jordan} already implies this result for G\CTNS{}, since it shows that any valid rational Gaussian kernel can be realized as \cref{eq_canonical_jordan}.
Below we provide an explicit prescription, reminiscent of Ostrogradsky's construction \cite{2015TheoremOstrogradsky}, which applies also to non-Gaussian \CTNS{}.

\begingroup
\let\fvind\hdvind
For simplicity, let us start with only the physical field ($\cbd=0$) and set
\begin{equation}
    \ctnsfunc(\ff) = \xint\: \ctnsfuncdens(\ff(\xxx),\nabla\ff(\xxx),\ldots,\nabla^{\dvord}\ff(\xxx))
    \ ,
\end{equation}
where $\dvord$ is the true maximal derivative order for $\ff$ (\idest{}, it cannot be reduced integrating by parts).
Introduce $\hdvind[-1]=\floor{\dvord/2}$ auxiliary-field pairs
\begin{math}
    \{(\dvaux,\dvmul)\}\indlims{\hdvind}
\end{math},
namely $\dvord$ new fields if $\dvord$ is even and $\dvord-1$ new fields if $\dvord$ is odd,
and define the first-order functional $\ctnsfunc'$
(the $\xxx$ dependence of the fields is implicit):
\begin{equation}\label{eq_higher_order_removal}
    \ctnsfunc'(\ff,\{\dvaux\},\{\dvmul\})
    = \xint \brackets[\bigg]{
        \ctnsfuncdens(\ff,\nabla\ff,\dvaux[0],\nabla\dvaux[0],\ldots,\dvaux[-1],\nabla\dvaux[-1])
        + \sum\indlims{\hdvind} \dvmul \brackets*{\dvaux-\nabla^2\dvaux<}
    }
    \ ,
\end{equation}
with the identification $\dvaux_0=\ff$.
The $\dvmul$ fields act as Lagrange multipliers, whose equations of motion enforce
\begin{equation}
    \dvaux = \nabla^2\dvaux<
    \quad\leadsto\quad
    \dvaux = \nabla^{2\hdvind}\!\ff
    \ ,
\end{equation}
establishing the equivalence of $\ctnsfunc'$ and $\ctnsfunc$.
When $\dvord$ is even, the last argument of $\ctnsfuncdens$ in \cref{eq_higher_order_removal} is absent and $\dvaux[-1]$ can be integrated as per \cref{eq_ctns_reduce} without spoiling locality, since
$\ctnsfunc'$ does not depend on its derivatives.
The result is a \CTNS{} with $\dvord-1$ auxiliaries for even and odd $\dvord$ alike.
\endgroup

For $\cbd$ initial auxiliaries, the above procedure is carried out independently for each $\ffv$, producing a \CTNS{} with
\begin{equation}\label{eq_higher_order_count}
    \cbd' = \sum\nolimits\indlims{\fvind} \dvordv - 1
    \ ,
\end{equation}
$\dvordv$ being the maximal derivative order of $\ffv$.
The counting simplifies to $\cbd' = \dvord(\cbd+1)-1$ for a uniform order $\dvordv=\dvord$.

By \cref{eq_higher_order_count} alone, one can conclude that higher-order Gaussian \CTNS{} are non-generic, in the sense that they correspond to non-generic first-order G\CTNS{} (at least if $\dvordv[0]>1$):
Contracting a G\CTNS{} with $\sum\indlims\vvind\dvordv=N$ gives a kernel $\kernel$ with denominator degree $N$; however, \cref{eq_higher_order_removal} rewrites it
as a first-order one whose $\kernel$'s denominator degree would generically be $\cbd'=N + (\dvordv[0]-1)$.
The $\dvordv[0]\mathbin-1$ degree gap corresponds to higher-order derivatives of the physical field, which do not enter the denominator because $\fpp$ is not integrated over in \cref{eq_ctns}.
To see why more explicitly, consider a $\cbd=0$
G\CTNS{} $\ctnsfunc$ with
\begin{equation}
    \ctnsfuncdens = \frac12 \ff P^{(\dvordv[0])}(-\nabla^2)\ff
    \ ;
\end{equation}
namely a Gaussian state with polynomial kernel $\kernel=P^{(\dvordv[0])}$ (\eg{}, a Lifshitz vacuum).
By \cref{eq_gctns_kernel_general}, the first-order G\CTNS{} $\ctnsfunc'$ corresponds (modulo field redefinitions) to a virtual block $\vctnsmat=\idmat+\nilmat\kk^2$, \idest{} the $\polenum\mathbin=0$ block of \cref{eq_canonical_jordan},
which does not contribute any denominator degree because $\nilmat$ is nilpotent and thus $\det\vctnsmat=1$.

\section{Continued fraction expansion of $\sqrt{1+\z}$}\label{sec_cf_convergence}

We collect below some convergence results for the continued fractions introduced in \cref{sec_gs}.
The central object of this \namecref{sec_cf_convergence} is the continued fraction representation of $\rootz=\sqrt{1+\z}$ with modified $\cfn$th convergent
\begin{equation}\label{eq_cf_square_root}
    \cfconv(\z,\w) =
    1 + \cfrac{\z}{
        2 + \cfrac{\z}{
            2 + \cfrac{\z}{
                \cfracdots+\cfrac{\z}{2+\w}}}}
    \ .
\end{equation}
Direct inspection reveals that the continued fractions $\padefreq$ \cref{eq_pade_cf} and $\evofreq$ \cref{eq_evo_W} are affinely equivalent to it --- \idest{}, their convergents are related by an affine transformation to \cref{eq_cf_square_root}:
\begin{subequations}
    \begin{align}
        \padefreq(\kk) & = \label{eq_cf_sqrt_pade}
        \mass \, \cfconv
        \parentheses*{
            \z\compeq\frac{\kk^2}{\mass^2}, \,
            \w\compeq0
        }
        \ ,                                        \\
        \evofreq(\kk)  & = \label{eq_cf_sqrt_evo}
        \frac{\evodia}{\evoeps} \, \cfconv[\evonum]
        \parentheses*{
            \z\compeq{-\frac{1}{\evodia^2}}, \,
            \w\compeq-\frac{\evodia_0}{\evodia}
        }
        + \frac{\evodia_0-\evodia}{\evoeps}
        \ ,
    \end{align}
\end{subequations}
where $\evodia$ and $\evodia_0$ are those in \cref{eq_trotter_coeffs}.

A closed-form expression for $\cfconv$
follows from standard techniques for periodic continued fractions \cite[Lemma~1.1]{Lorentzen2008ContinuedFractionsConvergence}.
Continuants (numerators and denominators of the convergents) satisfy the linear recurrence
\begin{equation}
    \cfrec_\cfn = 2 \cfrec_{\cfn-1} + \z \cfrec_{\cfn-2}
    \ ,
\end{equation}
with characteristic roots
\begin{math}
    \lambda_{\pm}  = 1 \pm \rootz
\end{math}.
Its general solution takes the form
\begin{equation}
    \cfrec_\cfn = c^{\vphantom{\cfn}}_+ \lambda_{+}^\cfn + c^{\vphantom{\cfn}}_- \lambda_{-}^\cfn
    \ ,
\end{equation}
where $c_\pm$ are set by the initial conditions:
\begin{itemize}
    \item for the numerator, $\cfrec_0 = 1$ and $\cfrec_1 = 2+\z$,
          corresponding to $c_{\pm}=\lambda_{\pm}/2$;
    \item for the denominator, $\cfrec_0 = 1$ and $\cfrec_1 = 2$,
          corresponding to $c_{\pm}=\pm\lambda_{\pm}/2\rootz$.
\end{itemize}
Using eq.~(1.2.6) of ref.~\cite{Lorentzen2008ContinuedFractionsConvergence} and simplifying, we obtain
\begin{equation}\label{eq_cf_sqrt_convergent}
    \cfconv(\z,\w)
    = \rootz \:
    \frac
    {1+\rootsratio[\w]\rootsratio^{\cfn}}
    {1-\rootsratio[\w]\rootsratio^{\cfn}}
    \ , \quad
    \rootsratio[\w]
    = \frac{\lambda_-+\w}{\lambda_++\w}
    = \frac{1+\w-\rootz}{1+\w+\rootz}
    \ .
\end{equation}
$\cfconv$ converges geometrically to $\rootz$ for all $\w\neq-\lambda_+$, as long as $\abs{\rootsratio}<1$, namely,
\begin{equation}\label{eq_cf_square_root_convergence}
    \lim_{\cfn\to\infty}\cfconv(\z,\w)
    =\cfconv[\infty](\z)
    =\rootz
    \qquad
    \forall\:
    \z>-1,\;
    1+\w+\rootz\ne0
    \ .
\end{equation}
Still, convergence is not uniform: it is faster when $\abs{\rootsratio}\ll1$ ($\z\to0$) and slows down as $\abs{\rootsratio}\to1^-$ ($\z\to-1^+,\infty$).

\subsection{Padé approximants \texorpdfstring{\cref{eq_pade_cf}}{\ref{eq_pade_cf}}}\label{sec_cf_convergence_pade}

\begingroup
\let\rootsratio\eta

Inserting \cref{eq_cf_sqrt_convergent} into \cref{eq_cf_sqrt_pade} gives
\begin{equation}
    \padefreq = \freq \frac{1+\rootsratio^{\cfn+1}}{1-\rootsratio^{\cfn+1}}
    \ , \qquad
    \rootsratio=\rootsratio(\z,0)=\frac{1-\rootz}{1+\rootz}
    \ ,
\end{equation}
which is the
$[\lceil\cfn/2\rceil/\lfloor\cfn/2\rfloor]$ Padé approximant of $\freq$ as a function of $\kk^2$, centered at $\kk=0$.
Indeed, multiplying the numerator and the denominator by $(1+\rootz)^{\cfn+1}$,
\begin{equation}
    \cfconv(\z,0)
    = \rootz\,
    \frac{(1+\rootz)^{\cfn+1}+(1-\rootz)^{\cfn+1}}
    {(1+\rootz)^{\cfn+1}-(1-\rootz)^{\cfn+1}}
    \ .
\end{equation}
The sum in the numerator contains only even powers of $\rootz$.
The difference in the denominator contains only odd powers, and cancelling one $\rootz$ factor against the prefactor makes $\cfconv(\z,0)$ rational in $\rootz^2=1+\z$,
with numerator and denominator degrees $\lceil\cfn/2\rceil$ and $\lfloor\cfn/2\rfloor$, respectively.
Moreover, for small $\z$,
\begin{equation}
    \rootsratio
    = -\frac{\z}{(1+\rootz)^2}
    = -\frac{\z}{4}+\order{\z^2}
    \ ;
\end{equation}
thus,
\begin{equation}
    \frac{\padefreq-\freq}{\mass}
    = \frac{2\rootz\rootsratio^{\cfn+1}}{1-\rootsratio^{\cfn+1}}
    = \order{\z^{\cfn+1}}
    \ ,
\end{equation}
which is the defining condition for the $[\lceil\cfn/2\rceil/\lfloor\cfn/2\rfloor]$ Padé approximant of $\freq$ at $\z=0$, equivalently $\kk=0$.
We note that a first-derivative tridiagonal G\CTNS{} with $\cbd$ virtual fields can also realize the maximal Padé approximant of degree $\kerdeg$, namely $\padefreq[2\cfn+1]$, by a contraction of the continued fraction \cref{eq_cf_square_root}, generated by the identity
\begin{equation*}
    \sqrt{1+\z}=1+\frac z2 -\cfrac{\z^2/4}{1+\z/2+\sqrt{1+\z}}
    \ .
\end{equation*}

The UV asymptotes \cref{eq_pade_asym} are recovered expanding for $\sqrt{\z}=\kk/\mass\to\infty$ at fixed $\cfn$ via
\begin{equation}
    \rootsratio^{n} \sim (-1)^{n}(1-2n/\sqrt{\z})
    \ .
\end{equation}
Since $\z=(\kk/\mass)^2\ge0$,
\cref{eq_cf_square_root_convergence} implies $\padefreq\to\freq$ for $\cfn\to\infty$, but convergence is slower for small masses and large momenta, as illustrated in \cref{fig_dispersion_convergence} (\cpageref{fig_dispersion_convergence}).
Indeed, the absolute relative error increases monotonically with $\z$:
\begin{equation}\label{eq_cf_err}
    \relerr=\frac{\padefreq-\freq}{\freq}
    = \frac{2\rootsratio^{\cfn+1}}{1-\rootsratio^{\cfn+1}} \underset{\cfn\to\infty}{\sim} 2\rootsratio^{\cfn+1}
    \ .
\end{equation}
Since $-1<\rootsratio\leq0$, $\relerr$ alternates sign with $\cfn$, as expected for a positive continued fraction \cite[Theorem~3.12]{Lorentzen2008ContinuedFractionsConvergence}.
The following expansions hold:
\begin{subequations}\label{eq_cf_err_expansions}
    \begin{equation}\label{eq_cf_err_ir}
        \abs{\relerr}\sim
        2(\kk/2\mass)^{2(\cfn+1)}
        \qquad
        (\kk/\mass\ll1,\text{ fixed }\cfn)
        \ ,
    \end{equation}
    and, using $\rootsratio=-e^{-2\operatorname{arcsinh}(\z^{-1/2})}$, to leading exponential order,
    \begin{equation}\label{eq_cf_err_uv}
        \abs{\relerr}\approx
        2e^{-2(\cfn+1)\mass/\kk}
        \qquad
        (1\ll\kk/\mass\ll\cfn)
        \ .
    \end{equation}
\end{subequations}

\endgroup

\subsection{Trotter approximants \texorpdfstring{\cref{eq_evo_W}}{\ref{eq_evo_W}}}\label{sec_cf_convergence_evo}
Inserting \cref{eq_cf_sqrt_convergent} into \cref{eq_cf_sqrt_evo} gives
\begin{equation}
    \evofreq    =
    \frac{\sqrt{\evodia^2-1}}{\evoeps}
    \frac{1+\rootsratio[\w]\rootsratio^{\evonum}}{1-\rootsratio[\w]\rootsratio^{\evonum}}
    + \frac{\evodia_0-\evodia}{\evoeps}
    \ .
\end{equation}
The limit in this case is slightly more subtle because the derivation changes depending on whether $\cfn=\evonum\to\infty$ is taken keeping $\evoeps$ or $\evoT$ fixed.
Fixing $\evoeps>0$, as in the main text,
\cref{eq_cf_square_root_convergence} applies straightforwardly and, recalling the values of $\evodia$ and $\evodia_1$ from \cref{eq_trotter_coeffs}, we conclude that the continued fraction \cref{eq_cf_sqrt_evo} converges to the result \cref{eq_evo_convergence} for all $\kk$ --- provided $\mass>0$, as expected.
Fixing $\evoT$, on the other hand, the fraction coefficients acquire a dependence on $\evonum$ through $\evoeps=\evoT/\evonum\to0$.
Consequently, \cref{eq_cf_square_root_convergence} (which assumes $\z,\w$ constant) does not apply.
Instead, recalling
$\evodia=1+\evoeps^2\freq^2/2$ (and thus $\sqrt{\evodia^2-1}\sim\evoeps\freq$),
and $\evodia_{0}-\evodia=\order{\evoeps^2}$,
\begin{equation*}
    \def\epsinv{\evonum}
    \rootsratio^{\epsinv}
    = \brackets*{\frac{\evodia-\sqrt{\evodia^2-1}}{\evodia+\sqrt{\evodia^2-1}}}^{\mathrlap{\epsinv}}
    \sim \brackets*{1-\frac{2\freq\evoT}{\evonum}}^{\mathrlap{\epsinv}}
    \to e^{-2\freq\evoT}
    \ , \quad
    \rootsratio[\w] = \frac{\evodia-\evodia_0-\sqrt{\evodia^2-1}}{\evodia-\evodia_0+\sqrt{\evodia^2-1}} \to -1
    \ ;
\end{equation*}
therefore
\begin{equation}\label{eq_cf_trotter_limit}
    \lim_{\evonum\to\infty}\evofreq[\evonum,\evoT/\evonum] = \freq \tanh(\freq\evoT)
    \ .
\end{equation}
The exact dispersion relation is recovered letting $\evoT\to\infty$,
while expanding $\tanh(\freq\evoT)\sim\freq\evoT$ for $\freq\evoT\ll1$ recovers the IR asymptote from \cref{sec_kg_dispersion}.

We conclude by noting that the IR expansion there ($X=\evoeps\freq$, $Y_{\evonum}=\evoeps\evofreq$),
\begin{equation*}
    Y_{\evonum+1}
    =
    1+(1-\trottercoeff)X^2 - \brackets*{1+\trottercoeff X^2+Y_{\evonum}}\inv
    \sim
    Y_{\evonum}+X^2
    \ ,
\end{equation*}
may seem justified as soon as
$Y_{\evonum'}\ll1$ for all $\evonum'<\evonum $,
\idest{},
\begin{math}
    \evoeps\evoT\freq^2\ll1
\end{math} ---
which implies $X=\evoeps\freq\ll1$, but is weaker than the actual condition $\evoT\freq\ll1$, found by matching to $\freq$ and confirmed by \cref{eq_cf_trotter_limit}.
The shorter time scale $\evoT\approx1/\freq$
emerges due to a secular accumulation of higher-order terms as  $\evonum$ grows:
Using the solution to the linearized recurrence to estimate the correction due to the neglected quadratic term in
\begin{math}
    [1+z]\inv = 1 - z + z^2 + \dots
\end{math},
one gets
\begin{equation*}
    \sum_{\evonum'=0}^{\evonum} (\trottercoeff X^2+Y_{\evonum'})^2
    = \sum_{\evonum'=0}^{\evonum} (\trottercoeff+\evonum')^2 X^4
    \sim \frac13\evonum^3 X^4
    \ .
\end{equation*}
Therefore,
\begin{math}
    Y_{\evonum} = \evonum X^2 \brackets*{1+\order{\evonum^2 X^2}}
\end{math},
and the IR expansion is self-consistent only when $\evonum X = \evoT\freq \ll1$.
In fact, to leading nontrivial order in $\evoeps\to0$ (no UV effects),
\begin{equation*}
    \def\evofreqT{\evofreq[\evoT/\evoeps,\evoeps]}%
    Y_{\evonum+1} - Y_{\evonum} = X^2 - Y_{\evonum}^2
    \quad\leadsto\quad
    \odv{\evofreqT}{\evoT} = \freq^2 - \parentheses*{\evofreqT}^2
    \quad\leadsto\quad
    \evofreqT = \freq \tanh(\freq\evoT)
    \ ,
\end{equation*}
matching the exact $\evoeps\to0$ result above.

\section{Zolotarev's rational approximation of $\sqrt{\z}$}\label{sec_zolotarev}
Writing
\begin{equation}
    \freq=\mass\sqrt{\z}
    \ , \qquad
    \z = 1+\kk^2/\mass^2
    \ , \qquad
    \zuv = 1+\pepsuv^2/\mass^2
    \ ,
\end{equation}
finding a G\CTNS{} approximation of Klein-Gordon's vacuum accurate up to the scale $\pepsuv$ with bond dimension $\cbd$ amounts to finding the best rational approximation $R(\z)$ of $\sqrt{\z}$ on $\zint$ of degree $(\cbd+1,\cbd)$.
Minimizing the uniform relative error $\unirelerr$,
\begin{equation}
    \unirelerr = \max_{\zint}\abs{\relerr(\z)}
    \ , \qquad
    \relerr(\z) = \frac{R(\z)}{\sqrt{\z}}-1
    \ ,
\end{equation}
leads to a classical minimax problem whose solution was given in closed form by Zolotarev in the 19th century \cite{Zolotarev1877ApplicationEllipticFunctions} and is now used in matrix square-root algorithms \cite{Gawlik2019ZolotarevIterationsMatrix} and overlap fermion lattice QCD computations \cite{Chiu2002NoteZolotarevOptimal}.
This appendix collects properties of Zolotarev's result used in the main text.

\begin{figure}
    \includegraphics{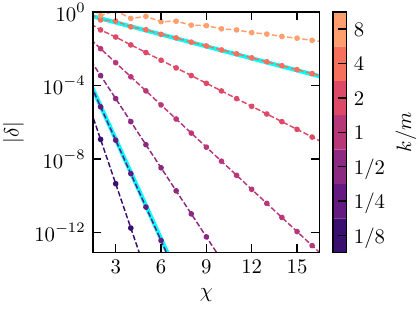}
    \quad
    \includegraphics{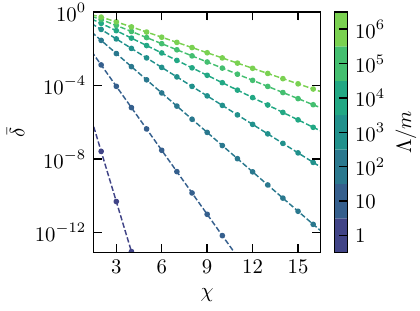}
    \caption{%
        Left: pointwise convergence of Klein-Gordon's Padé approximant $\padefreq$ in \cref{fig_dispersion_pade} for various momenta.
        The expansions \cref{eq_cf_err_expansions} (cyan) are accurate already for $\protect\kk/\mass\in\{1/4,4\}$.
        Right: uniform relative error $\unirelerr$ of the Zolotarev approximant \cref{eq_zolotarev} over $\protect\kk\in[0,\pepsuv]$, for various $\pepsuv/\mass$.
        Both errors decay exponentially in $\cbd$ but, at fixed target accuracy (horizontal cut), the required $\cbd$ grows linearly in the highest momentum to be resolved for Padé, \cref{eq_cf_err_uv}, and logarithmically for Zolotarev, \cref{eq_zolotarev_error}.
        Dashed lines are guides to the eye.
    }
    \label{fig_dispersion_convergence}
\end{figure}

Zolotarev's optimal approximant reads \cite{Gawlik2019ZolotarevIterationsMatrix,Chiu2002NoteZolotarevOptimal}
\begin{equation}\label{eq_zolotarev}
    \Rzol
    = d_0\,(\z+c_{2\cbd+1}) \prod_{l=1}^{\cbd} \frac{\z+c_{2l-1}}{\z+c_{2l}}
    \ , \qquad
    c_l = \frac{\sn^2(\vartheta_l;\kappa)}{\cn^2(\vartheta_l;\kappa)}
    \ , \qquad
    \vartheta_l = \frac{l\,K(\kappa)}{2\cbd+2}
    \ .
\end{equation}
Here $\sn$ and $\cn$ are Jacobi elliptic functions and $K$ is the complete elliptic integral of the first kind, with modulus
\begin{math}
    \kappa = \sqrt{1-1/\zuv}
\end{math}.
The normalization $d_0>0$ is fixed by centering the range of the relative error about zero:
\begin{equation}
    d_0 = \frac{2}{h_-+h_+}
    \ , \qquad
    \unirelerr = \frac{h_+-h_-}{h_++h_-}
    \ ,
\end{equation}
where
\begin{equation}
    1+\relerr(\z) = d_0\,h(\z)
    \ , \qquad
    h_- = \min_{\zint} h(\z)
    \ , \qquad
    h_+ = \max_{\zint} h(\z)
    \ .
\end{equation}
In this way, $\relerr$ equioscillates between $\pm\unirelerr$ at $2\cbd+3$ points in $\zint*$, characterizing $\Rzol$ as an optimal uniform approximant \cite{Chiu2002NoteZolotarevOptimal}.
Ref.~\cite[eqs.~(11) and~(27)]{Gawlik2019ZolotarevIterationsMatrix}
gives the asymptotically sharp bound
\begin{equation}\label{eq_zolotarev_error_bound}
    \unirelerr
    \leq 4\rho(\zuv)^{-2(\cbd+1)}
    \ , \qquad
    \rho(\zuv)
    = \exp*{\frac{\pi K(\zuv^{-1/2})}{K(\kappa)}}
    \ ;
\end{equation}
\begin{math}
    \unirelerr
    \sim
    4\rho(\zuv)^{-2(\cbd+1)}
\end{math}
as $\cbd\to\infty$ at $\zuv$ fixed (see \cref{fig_dispersion_convergence}).
Moreover,
\begin{equation}
    \log\rho(\zuv)
    \underset{\zuv\to\infty}{\sim}
    \pi^2/\log(16\zuv)
    \ .
\end{equation}
In the $1\ll\log\zuv\ll\cbd$ regime, combining these results gives, to leading exponential accuracy,
\begin{equation}\label{eq_zolotarev_error}
    \unirelerr
    \approx 4\exp*{-\frac{2\pi^2(\cbd+1)}{\log(16\zuv)}}
    \ .
\end{equation}
Since $\unirelerr$ is monotonic in both $\cbd$ and $\zuv$, given either of these and a target accuracy $\unirelerr$, the remaining parameter can be found by bisection.
In \cref{fig_dispersion_zolotarev,fig_entanglement,fig_entanglement_scaling}, we use this property to determine $\pepsuv$ given $\unirelerr$ and $\cbd$.

Since $0<c_1<c_2<\dots<c_{2\cbd+1}$, the approximant is positive for $\z\geq0$ and has simple poles at $\z=-c_{2l}$, $l=1,\dots,\cbd$, all with negative residues.
The increasing affine map $\z(\kk^2)$ preserves positivity and the signs of the residues, and sends the poles outside the physical region $\kk^2\geq0$.
Consequently, the kernel
\begin{equation}\label{eq_zolo_dispersions}
    \zolofreq(\kk) = \mass\Rzol(\z)
    \ ,
\end{equation}
defines a valid G\CTNS{} of the form \cref{eq_gctns_diagonal}, with $\cbd$ auxiliary fields and real couplings.
Zolotarev's construction also yields an optimal rational approximant of $\sqrt{\z}$ of degree $(\cbd,\cbd)$, and hence the optimal G\CTNS{} approximation of Klein-Gordon within the subclass with no physical-field gradient term in \cref{eq_gctns_W}, considered in ref.~\cite{Tilloy2019ContinuousTensorNetwork}, but its uniform relative error is strictly larger than that of \cref{eq_zolotarev} \cite{Gawlik2019ZolotarevIterationsMatrix}.

\input{manuscript.bbl}

\end{document}

%% file: manuscript.bbl
\providecommand{\href}[2]{#2}\begingroup\raggedright\endgroup